\documentclass[aoas]{imsart}
\RequirePackage{amsthm,amsmath,amsfonts,amssymb}
\RequirePackage[authoryear]{natbib}
\RequirePackage[colorlinks,citecolor=blue,urlcolor=blue]{hyperref}
\usepackage{graphicx}%
\usepackage{multirow}%
\usepackage{amsmath,amssymb,amsfonts}%
\usepackage{algorithm}
\usepackage{algorithmic} 
\usepackage{amsthm}%
\usepackage{mathrsfs}%
\usepackage{amsmath}
\usepackage{booktabs}
\usepackage{tikz}
\usepackage{pgfplots}
\usetikzlibrary{pgfplots.groupplots}
\usepackage{amsmath}
\usetikzlibrary{arrows.meta,positioning,calc}
\usetikzlibrary{arrows.meta,positioning}
\usepackage[title]{appendix}%
\usepackage{xcolor}%
\usepackage{textcomp}%
\usepackage{manyfoot}%
\usepackage{booktabs}%
\pgfplotsset{compat=1.18}   
\usetikzlibrary{arrows.meta, bending}
\usepackage{algorithm}%
\usepackage{listings}%
\usepackage{booktabs}%
\usepackage{amsmath,amsthm,latexsym,amssymb,enumerate,accents}

\newcommand{\E}{\mbox{$\mathbb{E}$}}

\startlocaldefs

\endlocaldefs

\begin{document}

\begin{frontmatter}
\title{Statistical Analysis of Network Collections Using Persistent Homology and Functional Data Analysis}

\runtitle{statistical analysis of collections of networks}

\begin{aug}

\author[A]{\fnms{Catherine}~\snm{Higgins}\ead[label=e1]{catherine.higgins@ucd.ie}\orcid{0009-0003-8718-9335}},
\author[B]{\fnms{Hulin}~\snm{Wu}\ead[label=e2]{hulin.wu@uth.tmc.edu}\orcid{0000-0002-5809-5407}}
\and
\author[A]{\fnms{Michelle}~\snm{Carey}\ead[label=e3]{michelle.carey@ucd.ie}\orcid{0000-0002-5603-4264}}

\address[A]{School of Mathematics and Statistics, University College Dublin,
Ireland\printead[presep={,\ }]{e1,e3}}

\address[B]{Department of Biostatistics and Data Science, The University of Texas
Health Science Center at Houston, Texas, United States of America\printead[presep={,\ }]{e2}}
\end{aug}

\begin{abstract}
Statistical analysis of collections of networks, where each network is treated as the primary unit of observation, is of growing importance across a wide range of application domains, including gene regulatory, social, neural, and financial networks. As networks consist of vertices and edges that do not naturally reside in Euclidean space, the direct application of conventional statistical methodologies, such as the computation of means and covariances, principal component analysis, and hypothesis testing, to samples of networks is not straightforward. A central challenge lies in defining meaningful measures of similarity or distance between networks of potentially varying sizes and structural types (e.g., directed, undirected, weighted or unweighted), particularly when no predefined node correspondence exists.

To address these challenges, we introduce a framework termed functional topological data analysis (funTDA), which integrates tools from functional data analysis and topological data analysis to facilitate exploratory data analysis and inference on samples of networks. The proposed framework enables the computation of summary statistics, including means and variances, and supports the application of principal component analysis and hypothesis testing to topological features extracted from network data. Through simulation studies involving networks with varying connectivity structures, we demonstrate the ability of funTDA to distinguish between distinct network configurations. The methodology is also illustrated through two real-data applications: networks constructed from pairwise word co-occurrences in novels by Jane Austen and Charles Dickens, and gene regulatory networks derived from gene expression measurements for seventeen individuals exposed to H3N2 influenza. In both applications, differences in network topology are assessed using principal component analysis and formal hypothesis testing.
\end{abstract}


\end{frontmatter}

\section{Introduction}

Network science has emerged as a powerful framework for modelling complex systems across numerous scientific domains, from social interactions and biological regulation to transportation and communication infrastructure. While early developments focused predominantly on the analysis of individual networks \citep{jackson2008social, networkbook, newman2010networks}, a growing class of scientific questions now demands methodologies capable of analysing collections of networks, where each network is itself a distinct observational unit. Examples include examining the temporal changes in functional brain networks, studying differences in gene regulatory networks across patients, or characterising variation in social networks across communities.

Extending classical statistical tools, such as hypothesis testing or principal component analysis, to network-valued data presents fundamental challenges. Networks are inherently combinatorial objects, defined by sets of vertices and the edges that encode relationships between them. For example, consider a gene regulatory network. Here, vertices correspond to genes, and edges represent regulatory interactions, that is the extent to which the expression of one gene influences the expression of another. This graph-based formulation places networks outside the familiar territory of Euclidean spaces that underlies conventional statistical methodology.  

Mathematically, a network can be represented as a graph, denoted by $G=(V,E)$, where the vertex set is $V=\lbrace v_1, v_2,\ldots, v_m\rbrace$, and the edge set is $E=\lbrace w_{ij}\rbrace$, with $w_{ij}$ denoting the weight of the edge connecting vertices $v_i$ and $v_j$. If no edge exists between the vertices, then $w_{ij}=0.$ A convenient algebraic representation of a graph is its  $m \times m$ adjacency matrix, denoted by $\textbf{A} = \begin{Bmatrix} \mathit{a_{ij}}\end{Bmatrix}$, where each element ${a_{ij}}$ is equal to ${w_{ij}}$ if ${i\rightarrow j \in E }$, and ${a_{ij}}=0$ otherwise. Networks may be classified as directed, in which case edges possess directionality and satisfies $a_{ij} \ne a_{ji}$, or undirected, where edges are symmetric such that $a_{ij} = a_{ji}$. Additionally, networks may be weighted, with edge values reflecting the strength of interactions, or unweighted, where edges are binary indicators of presence or absence. Networks may be directed or undirected, weighted or unweighted, depending on the nature of the relationships they represent.

The objective of this work is to conduct statistical inference on a collection of networks, \( G_{1}, \ldots, G_{N} \), considering each network \( G_{g} \) for \( g = 1, \ldots, N \) as an independent observational unit. By embedding these networks into a suitable framework, we aim to characterise their common structure, quantify variation across the collection, and ultimately enable statistical comparisons.

Traditional statistical analyses of network samples typically rely on standard statistical methods, such as hypothesis testing, applied to graph-theoretic features computed at the level of individual vertices or edges. Commonly used features include degree distributions, centrality measures, small-world characteristics, and clustering coefficients \citep{networkbook, yaverouglu2014revealing}. By mapping each network into a shared feature space, such approaches reduce the analysis to a comparison of vectors, enabling the use of standard Euclidean distance metrics. However, this dimensionality reduction often entails substantial loss of information about the global network structure, can be highly sensitive to variations in graph topology, and may fail to capture structural equivalence \citep{prvzulj2007biological, oldham2019consistency}. In particular, networks with similar feature values, such as matching centrality measures, may nonetheless differ substantially in the identities and relative positions of key vertices, thereby undermining meaningful comparison \citep{ma2020ensemble,cavallaro2024sensitivity}

When all networks share a common set of vertices, the statistical analysis reduces to the study of a collection of matrices. For example, \cite{ginestet2017hypothesis} analyzes functional neuroimaging data where the vertices correspond to anatomical regions, specifically, the same fifty cortical and subcortical regions across all subjects, and the edges represent functional connections between these regions. Similarly, \cite{severn2022manifold} examine networks of word co-occurrences constructed from literary texts, where the vertices of each network correspond to an identical vocabulary set. In both studies, networks are represented via their graph Laplacian matrices. For the $g$th network, the graph Laplacian matrix is defined as $\textbf{L}_{g}=\textbf{D}_{g}-\textbf{A}_{g}$, for $g=1,\ldots,N,$ where $\textbf{A}_{g}$ denotes the adjacency matrix and $\textbf{D}_{g}$ is the diagonal degree matrix given by, $\textbf{D}_{g}=\textrm{diag}(\textbf{A}_{g}\textbf{1}_{m}),$ with $\textbf{1}_{m},$ representing an $m$-dimensional vector of ones. For conducting statistical inference on the graph Laplacian manifold, \cite{ginestet2017hypothesis} employ the Euclidean metric while \cite{severn2022manifold} consider Euclidean, square-root Euclidean, and Procrustes size-and-shape metrics \citep{dryden2009non}. However, these approaches are restricted to undirected networks defined over a common vertex set. Moreover, estimating  the covariance matrix of $m \times m$ graph Laplacians requires determining $\frac{m(m-1)(m^2-m+2)}{8}$ parameters. The accurate estimation and inversion of such a high-dimensional covariance matrix severely limits the practical size of networks that can be efficiently analysed using this framework.

In many real-world applications, it is essential to accommodate variability in both the number of vertices $m_g$ across networks, $g=1,\ldots,N,$ and their corresponding adjacency matrices $\textbf{A}_{g}$. Networks with non-common vertex sets arise frequently across a wide range of disciplines. For example, in gene regulatory network studies, different individuals may activate distinct subsets of genes, represented as vertices in each network, in response to a disease or infection. These activated genes may exhibit different interaction patterns, resulting in heterogeneous adjacency matrices across subjects. In such settings, the ordering of vertices within each network is arbitrary, and this lack of correspondence substantially complicates subsequent statistical analysis.

To address this \cite{jain2009structure} developed a framework in which adjacency matrices are represented within a quotient space of a Euclidean vector space under permutation actions, thereby formalizing graph equivalence up to vertex relabeling. Let the collection of adjacency matrices be denoted $\mathcal{A}=\{\textbf{A}_{1},\ldots,\textbf{A}_{N}\}$. Two matrices $\textbf{A}_{g} \in \mathcal{A}$ and $\textbf{A}^{'}_{g} \in \mathcal{A}$ are considered equivalent if there exists a permutation matrix $\textbf{P} \in \mathcal{T}$ such that $\textbf{P}\textbf{A}_{g}\textbf{P}^T = \textbf{A}^{'}_{g},$ where \( \mathcal{T} \) denotes the set of all possible \( m \times m \) permutation matrices. A permutation matrix is characterised by having exactly one entry equal to $1$ in each row and column, with all remaining elements equal to $0$. The resulting quotient space $\mathcal{T}_{\mathcal{A}} = \mathcal{A} / \mathcal{T}$ represents the set of all equivalence classes of adjacency matrices under permutation.

Within this quotient space, a graph alignment metric is defined to measure dissimilarity between two adjacency matrices by minimizing over all possible vertex relabelings, 
\begin{equation}\label{Dis_Q}
d_{g}([\textbf{A}_{g}],[\textbf{A}^{'}_{g}])= \min\limits_{\textbf{P} \in \mathcal{T}} d_{a}(\textbf{A}_{g},\textbf{P}\textbf{A}^{}_{g}\textbf{P}^T).
\end{equation}
In their original formulation, \cite{jain2009structure} proposed the Frobenius norm as the distance $d_{a}$. However, computing the optimal permutation matrix that attains this minimum is an NP-hard problem, which severely restricts the size of networks that can be analysed in practice. Building on this framework, \cite{jain2016statistical} and \cite{kolaczyk2020averages} introduced the Fréchet mean $\bar{\textbf{M}}$ of the sample graphs
defined by the minimum of the total squared Frobenius distance
\(
\sum_{g=1}^{N}\min \limits_{\textbf{P}\in \mathcal{T}}\|\textbf{P}\textbf{A}^{}_{g}\textbf{P}^T-\textbf{M}\|_F^{2}.
\) 
Subsequent developments extended this quotient-space perspective to inferential tasks beyond averaging. In particular,
\cite{guo2021quotient} implemented PCA for undirected/directed and weighted/unweighted networks by performing Euclidean PCA on differences between the Fréchet mean and the optimally aligned graphs. More recently, \cite{calissano2023populations} proposed the align-all-and-compute approach, that results in geodesic PCA for unlabeled directed/undirected and weighted/unweighted networks. However, the computational expense associated with the alignment of the graphs limits the size and number of networks that can be analysed. These limitations motivate the need for representations that capture global network structure, accommodate varying vertex sets, and permit scalable statistical inference. In this work, we propose a new framework, functional topological data analysis (funTDA), designed to meet these requirements.

Topological data analysis \citep{carlsson2009topology},  
specifically persistent homology,  offers a technique for identifying topological properties, such as cycles, across multiple scales \citep{edelsbrunner2002topological,zomorodian2005computing}, emphasizing the structural importance of the properties enduring over extended periods. By tracking the emergence and persistence of these features across resolutions, this approach captures global structural characteristics of a network rather than focusing solely on local motifs. Persistent homology can be summarized in several equivalent representations, most commonly persistence diagrams \citep{edelsbrunner2002topological}. To date, much of the statistical analysis of persistence diagrams has relied on distance-based comparisons, such as the Wasserstein and bottleneck distances \citep{mileyko2011probability,turner2014frechet}. However, computing basic statistical summaries, such as averages, of persistence diagrams remains non-trivial; see \cite{bubenik2017persistence} for a detailed discussion.

Persistence landscapes \citep{bubenik2015statistical} address this by providing a functional representation of persistent homology in a Hilbert space, thereby facilitating statistical analysis \citep{bubenik2017persistence, berry2020functional}. This representation enables a natural transformation from a collection of networks to persistence diagrams, and subsequently to persistence landscape functions, where each function encodes the persistent homology features of a single network. By representing persistent homology as functions, we can apply tools from functional data analysis \citep{FDAB, ramsay2009functional, kokoszka2017introduction}, which offer a framework for analyzing samples of functions. This allows us to quantify variability across networks, assess group differences, and perform formal statistical inference on network-derived data.

The key contribution of this paper is the proposal of funTDA, a framework that integrates persistent homology with functional data analysis to enable the statistical analysis of samples of networks. A principal advantage of topological data analysis over traditional graph-theoretic approaches is its ability to capture global network structure. Unlike methods that reduce networks to node or edge level summary statistics, it does not require node correspondence or equal network size, as is commonly assumed in Laplacian-based approaches. Moreover, it avoids the computational burden associated with graph matching, as encountered in quotient-space methods. Finally, the proposed framework is applicable to both weighted and unweighted networks, as well as to directed and undirected networks. The implementation of the proposed approach is available at \url{https://github.com/CatherineH1/funTDA}.

The remainder of this paper is organized as follows. Section~\ref{sec2} introduces the proposed approach for the statistical analysis of network-based data using funTDA. Section~\ref{sec3} presents a series of simulation studies designed to evaluate the performance of the proposed approach. Networks are generated using both Erdős–Rényi random graph models, in which edge presence follows a Bernoulli distribution and edge weights are drawn from a Beta distribution, and fully connected weighted stochastic block models. Where feasible, the proposed method is also compared with established approaches, specifically the quotient space framework of \cite{guo2021quotient} and the graph Laplacian method of \cite{severn2022manifold}, using both Erdős–Rényi networks and in Section \ref{literarycomparison} the literary networks considered in \cite{severn2022manifold}. Section \ref{sec4} is dedicated to the analysis of gene regulatory networks, which illustrate the genetic responses of nine symptomatic and eight asymptomatic subjects to intranasal H3N2 influenza exposure. The primary objective is to identify statistically significant differences between the networks representing symptomatic and asymptomatic subjects. Section \ref{sec5} concludes with final remarks.

\section{Statistical Analysis of Collections of Networks}\label{sec2}

This section describes the proposed funTDA approach for performing statistical analysis of network samples by combining persistent homology with functional data analysis techniques. Our presentation is limited to the essential elements of persistent homology that are required for understanding this methodology. For thorough introductions and formal definitions associated with persistent homology, please refer to \cite{edelsbrunner2002topological, zomorodian2005computing, carlsson2009topology} and \cite{edelsbrunner2010computational}. For details on computational implementation of persistent homology, please refer to \cite{lutgehetmann2020computing} and \cite{tauzin2021giotto}.

\subsection{Persistent Homology}\label{PH}

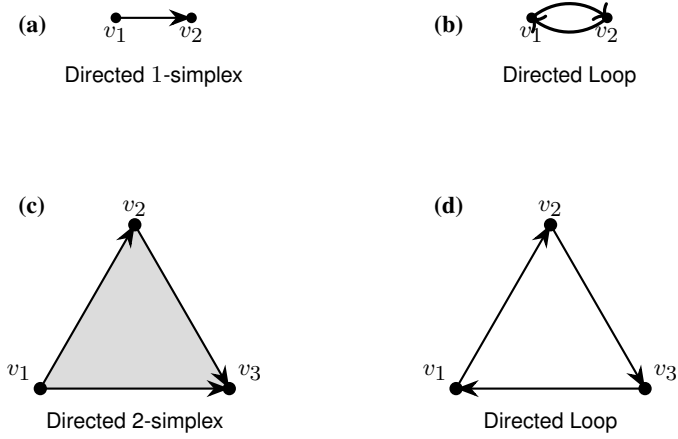
\begin{figure}[h!]
\centering
\begin{tikzpicture}[
        font=\sffamily,
        vertex/.style   = {circle,fill=black,inner sep=2pt},
        edge/.style     = {thick},
        dedge/.style    = {thick,-{Stealth[length=3mm]}},
        simplex/.style  = {fill=gray!45,opacity=.6}
    ]

\begin{scope}[shift={(0,0)}]

\node[anchor=west,font=\bfseries] at (-0.4,0.3) {(a)};
 \coordinate (a1) at (1.0,0.4);
  \coordinate (a2) at (2.0,0.4);
  \fill[vertex] (a1) circle (2pt) (a2) circle (2pt);
  \draw[dedge] (a1)--(a2);
  \node[below]  at (a1) {$v_{1}$};
  \node[below] at (a2) {$v_{2}$};
 \node[below=15pt] at ($(a1)!0.5!(a2)$) {Directed 
 $1$‑simplex};

 \node[anchor=west,font=\bfseries] at (5.1,0.3) {(b)};
 \coordinate (a1) at (6.5,0.4);
  \coordinate (a2) at (7.5,0.4);
  \fill[vertex] (a1) circle (2pt) (a2) circle (2pt);
  \draw[->, line width=1.2pt, bend left=40]  (a1) to (a2);
  \draw[->, line width=1.2pt, bend left=40]  (a2) to (a1);
  \node[below]  at (a1) {$v_{1}$};
  \node[below] at (a2) {$v_{2}$};
 \node[below=15pt] at ($(a1)!0.5!(a2)$) {Directed 
 Loop};

\end{scope}

\begin{scope}[shift={(0,-4.5)}]
  
  \node[anchor=west,font=\bfseries] at (-0.4,2.4) {(c)};

  \coordinate (dA) at (0,0);
  \coordinate (dC) at (2.5,0);
  \coordinate (dB) at ($(dA)!0.5!(dC)+(0,2.165)$);

  \fill[simplex] (dA)--(dB)--(dC)--cycle;
  \draw[dedge] (dA)--(dB);
  \draw[dedge] (dB)--(dC);
  \draw[dedge]  (dA)--(dC);  
  \fill[vertex] (dA) circle (2.5pt) (dB) circle (2.5pt) (dC) circle (2.5pt);

  \node[above left]  at (dA) {$v_{1}$};
  \node[above]       at (dB) {$v_{2}$};
  \node[above right] at (dC) {$v_{3}$};

  \node[below=6pt] at ($(dA)!0.5!(dC)$) {Directed 2‑simplex};
\end{scope}

\begin{scope}[shift={(5.5,-4.5)}]
  \node[anchor=west,font=\bfseries] at (-0.4,2.4) {(d)};

  \coordinate (eA) at (0,0);
  \coordinate (eC) at (2.5,0);
  \coordinate (eB) at ($(eA)!0.5!(eC)+(0,2.165)$);

  \draw[dedge] (eA)--(eB);
  \draw[dedge] (eB)--(eC);
  \draw[dedge] (eC)--(eA);

  \fill[vertex] (eA) circle (2.5pt) (eB) circle (2.5pt) (eC) circle (2.5pt);

  \node[above left]  at (eA) {$v_{1}$};
  \node[above]       at (eB) {$v_{2}$};
  \node[above right] at (eC) {$v_{3}$};

  \node[below=6pt] at ($(eA)!0.5!(eC)$) {Directed Loop};
\end{scope}

\end{tikzpicture}
 \caption{Examples of simplices and loops. (a) A directed 1-simplex. (b) A directed loop on two vertices. (c) A directed 2-simplex. (d) A directed loop on three vertices. \label{Figsimplex}}
 \end{figure}

Topological data analysis characterizes the structural properties that govern connectivity among data points in high-dimensional spaces. In the network setting, vertices are represented as points and edges as line segments or curves connecting them.  First-order homological attributes, $H_1$, commonly referred to as loops, correspond to circular structures that indicate the presence of directed cycles in the network. For example, Figure \ref{Figsimplex} (b) illustrates a loop formed by two vertices with directed edges $v_1 \rightarrow v_2$ and $v_2 \rightarrow v_1$ while Figure \ref{Figsimplex} (d) shows a loop involving three vertices connected by directed edges $v_1 \rightarrow v_2$, $v_2 \rightarrow v_3$, and $v_3 \rightarrow v_1$. 

In contrast, simplices represent higher-order interactions that are acyclic. A $k$-simplex of the directed graph $G$ is a collection of $k+1$-vertices $(v_{1}, ... , v_{k+1})$ of $G$ with the property that there is an edge $v_{i} \rightarrow v_{j}$ whenever $i < j$. Figure \ref{Figsimplex} (a) depicts a 1-simplex consisting of vertices $v_1$ and $v_2$, connected by the directed edge $v_1 \rightarrow v_2$. Figure \ref{Figsimplex} (c) shows a 2-simplex formed by three vertices with directed edges $v_1 \rightarrow v_2$, $v_1 \rightarrow v_3$, and $v_2 \rightarrow v_3$. This construction naturally extends to higher-order simplices, capturing relationships where connectivity flows uniformly from a source vertex to a sink.

From the perspective of persistent homology, loops correspond to nontrivial homology classes whose emergence and persistence across scales capture stable cyclic structure in the network, while simplices progressively fill in these loops and determine when such features disappear from the filtration. Persistent homology summarizes this evolution through birth–death pairs $(b_j, d_j)$, for $j =1,\cdots,|D|$, wherein $b_j$ specifies the scale at which a feature emerges (e.g., the formation of a loop), $d_j$ indicates the scale at which the feature ceases to be distinct (e.g., when the loop is absorbed into a higher-dimensional simplex), and $|D|$ is the total number of detected features. The persistence, $d_j-b_j$ thus quantifies the lifespan of each feature and serves as a measure of its structural significance.

To construct the filtration, we first compute the geodesic distance matrix $\boldsymbol{\Gamma }_{g} = \begin{Bmatrix} \mathit{\gamma_{ij}^{(g)}}\end{Bmatrix}$, such that the element ${\gamma_{ij}^{(g)}}$ is the length of the shortest directed path from vertex ${i\rightarrow j}$, if a path exists. For weighted graphs, this corresponds to the path minimizing the sum of edge weights, whereas for unweighted graphs it corresponds to the path with the fewest edges. The filtration parameter is then defined as the ordered set of distinct values in $\boldsymbol{\Gamma }_{g}$, i.e,  $\boldsymbol{\epsilon}_{g}=\{\min(\boldsymbol{\Gamma }_{g}),\ldots,\max(\boldsymbol{\Gamma }_{g})\}$. For each filtration value $\epsilon_{g,l}$, for $l=1,\ldots,L$ we construct the subgraph $G_{g,\epsilon_{g,l}}$ that retains the original vertex set but includes only those edges whose geodesic distance satisfies $\gamma_{ij}^{(g)} \leq \epsilon_{g,l}$. As $\epsilon_{g,l}$ increases, we track the birth and death of topological features across the filtration. 

\begin{figure}[h!]
 \centering
 {{\includegraphics[width = 14cm]{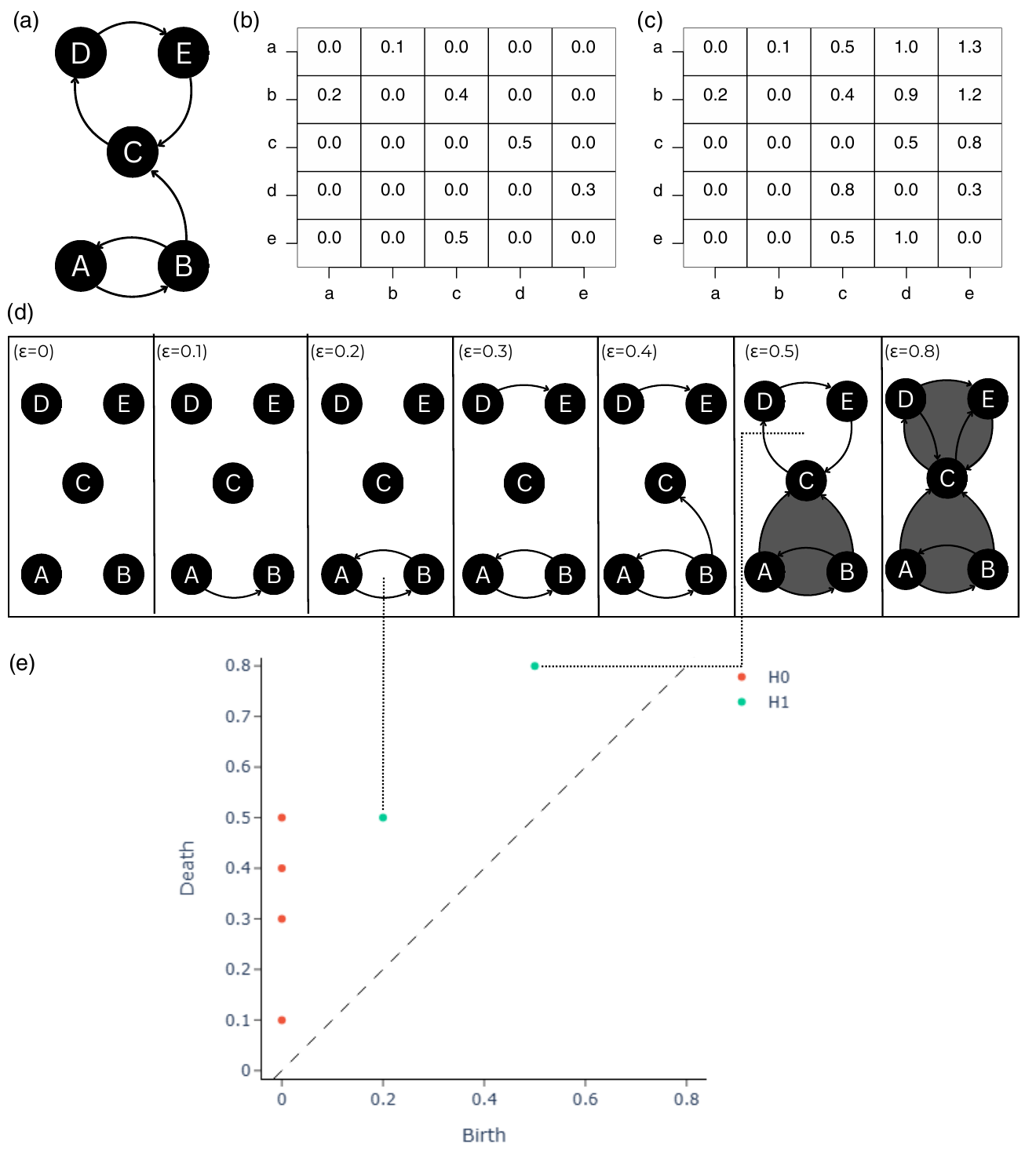} }}%
 
 \caption{Illustrative example of persistent homology for networks. (a) Network with two loops. (b) Adjacency matrix. (c) Geodesic Distance matrix. (d) Persistent homology filtration process. (e) Persistence diagram. \label{Fig1}}
 \end{figure}

As an illustrative case, Figure \ref{Fig1} (a) shows an example of a weighted directed network containing two loops, its corresponding adjacency matrix is shown in panel (b). The geodesic distance matrix where each entry $\gamma_{i,j}$ represents the length of the shortest directed path from vertex $i$ to vertex $j$ is shown in panel (c). The filtration parameter is defined over the range from the minimum distance of $0$ to the maximum distance $1.3$, yielding the ordered set $\boldsymbol{\epsilon}=\{0,\ldots,1.3\}$. 

Figure \ref{Fig1} (d) shows the sequence of networks obtained at selected filtration values $\{0,0.1,0.2,0.3,0.4,0.5,0.8\}$. At each filtration level $\epsilon_{l},$ an edge is included between a pair of vertices whenever the corresponding geodesic distance satisfies $\mathit{\gamma_{ij}} \leq \epsilon_{l}$. As $\epsilon_{l}$ increases, edges are added and each resulting graph is a subgraph of the subsequent one, forming a nested filtration. The appearance of a loop at a given filtration level is recorded as its birth time $b_j$, while the filtration value at which the loop is filled in by higher-dimensional simplices is recorded as its death time $d_j$. In Figure \ref{Fig1} (a), the network displays two loops, with the initial loop commencing at $\epsilon_{3}=0.2$. As the filtration value increases, more edges appear, and the loop dies once it is absorbed into a 2-simplex at \( \epsilon_{6} = 0.5 \). The second loop begins at \( \epsilon_{6} = 0.5 \) and persists until \( \epsilon_{9} = 0.8 \), where it is absorbed into two 2-simplexes.

\subsubsection{Persistence Diagrams}

The birth death pairs, $(b_j,d_j)$, can then be represented graphically using a persistence diagram \citep{edelsbrunner2002topological}. A persistence diagram is a multiset of points in the plane, where each point $(b_j,d_j)$, represents a topological feature born at filtration value $b_j$ and dying at $d_j$. As deaths occur subsequent to births, all points are situated above the diagonal line. Figure \ref{Fig1} (e) shows the resulting persistence diagram for the network in Figure \ref{Fig1} (a). The two green points represent the two loops identified in the network, with coordinates corresponding to their birth and death times. The red points represent $H_0$ features (connected components), which are discussed in Appendix~\ref{H_0}.

As the primary interest of this work lies in cyclic structure, we focus exclusively on  \(H_1\) loop features. This focus is further
motivated by the fact that \(H_0\) features can be readily characterized using standard results from spectral graph theory, that is the multiplicity of the zero eigenvalue of the graph Laplacian \citep{bollobas1998modern}. The persistence diagram associated with loop features can therefore be expressed as the set of birth--death pairs
 $
 \textbf{P}_{g} = \{(b_j,d_j): j =1,\cdots,|D|\},
 $
 where \(b_j\) and \(d_j\) denote the birth and death times of the \(j\)th loop, respectively, and \(|D|\) is the number of $H_1$ features with positive persistence (i.e., points strictly above the diagonal) in the $g^{th}$ diagram. For the example shown in Figure~\ref{Fig1}, this set is $\mathbf{P} = \{(0.2,0.5), (0.5,0.8)\}$. The distance of a point from the diagonal reflects the persistence \(d_j - b_j\) of the corresponding feature.

In the statistical analysis of network samples, each network gives rise to a persistence diagram, resulting in a collection of diagrams $\textbf{P}_1,\dots, \textbf{P}_N.$ Direct statistical analysis of persistence diagrams is challenging because they may differ in both the number and configuration of topological features, and the topological features within a diagram are not independent, as all features arise from the same underlying filtration process. The appearance or disappearance of one feature is inherently linked to the structure of the data and the sequence in which simplices are added. As a consequence, basic algebraic operations such as addition, averaging, or variance computation are not well defined, making it difficult to construct standard statistical summaries, even when the diagrams arise from a common underlying distribution \citep{bubenik2017persistence,berry2020functional}. One approach to summarizing multiple persistence diagrams is the Fréchet mean, as proposed by \cite{turner2014frechet}. However, the Fréchet mean in this setting can be unstable and is not guaranteed to be unique. Although a probabilistic framework ensuring a unique Fréchet mean has been developed \citep{munch2015probabilistic}, the associated computational cost limits its practical applicability. These challenges motivate the use of functional representations of persistent homology, such as persistence landscapes, which embed topological features in a Hilbert space and are therefore more amenable to statistical analysis.

\subsubsection{Persistence Landscapes}

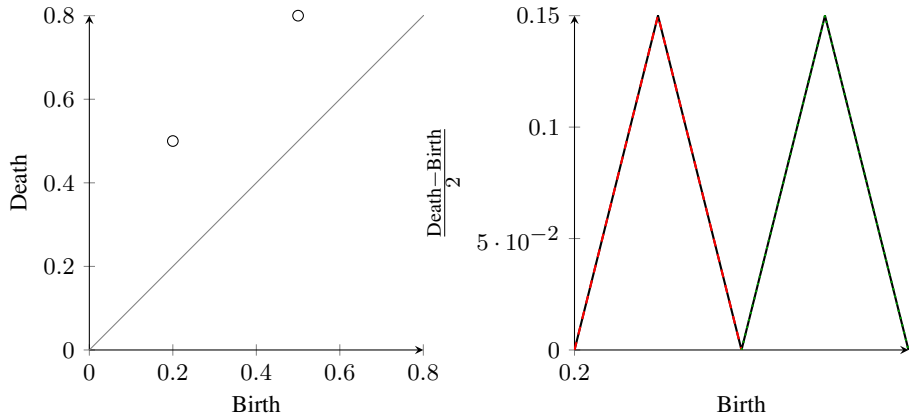
\begin{figure}[h!]
 \centering
\begin{tikzpicture}

\begin{groupplot}[
  group style={
    group size=2 by 1,
    horizontal sep=2.0cm,          
  },
  width=6cm,height=6cm,
]

\nextgroupplot[
  xlabel={Birth},
  ylabel={Death},
  xmin=0,xmax=0.8,
  ymin=0,ymax=0.8,
  axis equal image,
  axis lines=left,
  xtick={0,0.2,...,0.8},
  ytick={0,0.2,...,0.8},
]
\addplot[gray,thin,domain=0:1,samples=2] {x};

\addplot[
  only marks,
  mark=o,
  mark size=2pt,
] coordinates {
  (0.2,0.5)
  (0.5,0.8)
};

\nextgroupplot[
  xlabel={Birth},
  ylabel={$\frac{\textrm{Death}-\textrm{Birth}}{2}$},
  xmin=0.2,xmax=0.8,
  ymin=0,ymax=0.15,
  axis lines=left,
  xtick={0.2,...,0.8},
  ytick={0,0.05,0.1,0.15},
]

\addplot[
  thick,
  mark=none,
] coordinates {
  (0.2,0.00) (0.35,0.15) (0.5,0.00)
  (0.5,0.00) (0.65,0.15) (0.8,0.00)
  
};

\addplot[
  thick,
  red,
  dashed,
  mark=none,
] coordinates {
   (0.2,0.00) (0.35,0.15) (0.5,0.00)

};

\addplot[
  thick,
  green!60!black,
  dotted,
  mark=none,
] coordinates {
   (0.5,0.00) (0.65,0.15) (0.8,0.00)

};

\end{groupplot}
\end{tikzpicture}
\caption{An example of persistence landscapes. (Left) Persistence diagram illustrating two loop features. (Right) The corresponding first persistence landscape (black) is obtained as the pointwise maximum of the individual triangular functions (red and green) associated with each loop feature. \label{Fig_PDa}}
\end{figure}

The persistence landscape \citep{bubenik2015statistical} transforms a persistence diagram into a sequence of real-valued functions 
\(\{F_{g,(k)}\}_{k=1}^{K}\), with $K \leq |D|$ each defined on a common domain. This functional representation is constructed as follows. Consider the birth-death pairs $(b_j,d_j)$ in $\textbf{P}_{g} \in \mathcal{P}$ and define the piecewise linear function
\begin{equation}
 f_{(b_j,d_j)}(x) =
  \begin{cases}
   0 & \text{if } x \not \in (b_j,d_j) \\
   x - b_j & \text{if } x \in (b_j , \frac{b_j+d_j}{2}] \\
   -x + d_j       & \text{if } x \in (\frac{b_j+d_j}{2} , d_j), 
  \end{cases}
  \label{eq:basicLand}
\end{equation}
for $j=1,\ldots,|D|$ \citep{bubenik2015statistical,bubenik2017persistence}. As illustrated in Figure~\ref{Fig_PDa}, this transformation can be understood as mapping each point ($b_j,d_j$) in the persistence diagram to the apex of a triangle located at $(\frac{b_j+d_j}{2},\frac{d_j-b_j}{2})$. The height of the triangle thus corresponds to half the persistence of the feature, with more persistent features producing taller triangles. Each function \(F_{g,(k)} :
\mathbb{R} \to \mathbb{R}\) is obtained by ordering, at each location \(x\), the heights of the triangular functions, $\{ f_{(b_j,d_j)}(x) \}_{j=1}^{|D|}$ and selecting the \(k\)th largest value. The highest level $k=1$ corresponds to features with maximal persistence, which are typically interpreted as capturing the most significant underlying topological structure, while lower levels increasingly reflect transient or noise-induced phenomena. This representation embeds topological features into a common functional space.
Figure~\ref{Fig_PDa} illustrates this construction. The left panel shows the
persistence diagram consisting of the birth--death pairs corresponding to the two loop features in the example in Figure \ref{Fig1}. The right panel displays the associated first persistence landscape \(\mathbb{F}_{(1)}\) (shown in black), which is obtained as the pointwise maximum of the individual red and green triangular functions, $\{f_{(b_1,d_1)}(x),f_{(b_2,d_2)}(x)\},$ defined in \ref{eq:basicLand}. In practice, the collection of functions defined in \ref{eq:basicLand} is typically truncated to a finite number of layers $K<|D|$, retaining only the $K$ largest values at each point $\textbf{x}$. This truncation reflects the fact that only a finite subset of features contributes nontrivially to the persistence landscape at any given location. 

\subsection{Statistical Analysis}

We assume that the $N$ persistence diagrams $\mathbf{P}_1,\ldots,\mathbf{P}_N$ are independent and identically distributed draws from an unknown population distribution $\mathcal{P}$ \citep{berry2020functional}. 
The persistence landscape transformation provides a functional representation of each diagram. 
For a fixed landscape layer $k$, let $F_g(x)$ denote the true persistence landscape function for sample $g$. 
In practice, we do not observe $F_g(x)$ directly but rather evaluate it at a discrete set of points $\mathbf{x} = (x_1,\ldots,x_Q)^\top$ spanning the interval $I = [\min(\mathbf{b}), \max(\mathbf{d})]$, where $\mathbf{b}$ and $\mathbf{d}$ are the collections of birth and death times across all samples.

These evaluations are subject to observational error arising from multiple sources: the finite precision of geodesic distance computations, discretization of the filtration parameter, and numerical approximation in the landscape construction. 
We therefore model the observed values as
\begin{equation}\label{model:FDA}
\mathbf{F}_g^{\text{obs}} = \mathbf{F}_g + \boldsymbol{\varepsilon}_g, \qquad g = 1,\ldots,N,
\end{equation}
where $\mathbf{F}_g = F_g(\mathbf{x})$ is the $Q \times 1$ vector of true landscape values at the evaluation points, and $\boldsymbol{\varepsilon}_g$ is a mean-zero error vector. 
Following standard practice in functional data analysis \citep{FDAB}, we assume the errors are independent across samples and, for simplicity, that they are homoscedastic.

To represent the true underlying functions, we expand each $F_g(x)$ in a B‑spline basis:
\begin{equation}
F_g(x) = \sum_{l=1}^L \phi_l(x) c_{l,g},
\end{equation}
where $\{\phi_l(x)\}_{l=1}^L$ are second-order B‑spline basis functions and $c_{l,g}$ are the associated basis coefficients. 
Second-order B‑splines are chosen because they provide a flexible yet parsimonious representation of the piecewise linear structure of persistence landscapes, with the number of basis functions $L$ as $Q-1$.

Let $\boldsymbol{\Phi}$ denote the $Q \times L$ design matrix with entries $\phi_l(x_q)$. 
In matrix form, the model in \ref{model:FDA} becomes
$\mathbf{F}_g^{\text{obs}} = \boldsymbol{\Phi} \mathbf{c}_g + \boldsymbol{\varepsilon}_g$, for $g = 1,\ldots,N$,
where $\mathbf{c}_g = (c_{1,g},\ldots,c_{L,g})^\top$. 
The basis coefficients are estimated by ordinary least squares $
\hat{\mathbf{c}}_g = (\boldsymbol{\Phi}^\top \boldsymbol{\Phi})^{-1} \boldsymbol{\Phi}^\top \mathbf{F}_g^{\text{obs}}$.

Under the homoscedastic error assumption, the covariance matrix of the estimated coefficients is
$\operatorname{Var}(\hat{\mathbf{c}}_g) = \sigma^2 (\boldsymbol{\Phi}^\top \boldsymbol{\Phi})^{-1}$,
where $\sigma^2$ is estimated by
$\hat{\sigma}^2 = \frac{1}{N(Q-L)} \sum_{g=1}^N \| \mathbf{F}_g^{\text{obs}} - \boldsymbol{\Phi} \hat{\mathbf{c}}_g \|^2.$ The fitted persistence landscape for sample $g$, evaluated at the grid points, is $\hat{\mathbf{F}}_g = \boldsymbol{\Phi} \hat{\mathbf{c}}_g$, with pointwise covariance matrix $\operatorname{Var}(\hat{\mathbf{F}}_g) = \hat{\sigma}^2 \boldsymbol{\Phi} (\boldsymbol{\Phi}^\top \boldsymbol{\Phi})^{-1} \boldsymbol{\Phi}^\top$. Further details on spline smoothing within a functional data analysis context can be found in \citet{FDAB}.

\subsubsection{Mean and Variance}

The population mean persistence landscapes across the $N$ samples is given by
$
\mathbb{E}(F(x)),
$
 for $x\in\mathcal{I}$. A pointwise sample estimator is 
$
\bar{F}(x) = \frac{1}{N}\sum_{g=1}^N \hat{F}_g(x)=\sum_{l=1}^{L}\phi_{l}(x)\bar{c_{l}},
$
where $\bar{c_{l}} = \frac{1}{N}\sum_{g=1}^N \hat{c}_{l,g}$ is the average of the estimated $\ell$th basis coefficient across the $N$ samples. Convergence of $\bar{F}(x)$ towards $\mathbb{E}(F(x))$ was established in \cite{chazal2014stochastic}. 

The variance–covariance function quantifies the variability of persistence landscapes between samples and is estimated by
\begin{equation}\label{cov}
\textrm{Cov}(x,y) =  \frac{1}{N-1}
        \sum_{g=1}^{N}
        \bigl(\hat{F}_{g}(x)-\bar{F}(x)\bigr)
        \bigl(\hat{F}_{g}(y)-\bar{F}(y)\bigr),
   \qquad x,y\in\mathcal I .
\end{equation}

\subsubsection{Functional Principal Components Analysis (FPCA)}
 FPCA is primarily utilized for identifying the dominant modes of variation in a sample of functions, i.e., $\hat{F}_1,\ldots,\hat{F}_N$ typically after subtracting
the mean from each observation \citep{ramsay1991some,FDAB}. Under mild regularity conditions, there exists an orthonormal sequence of eigenfunctions
$\{\psi_{p}\}_{p=1}^{\infty}\subset L^{2}(\mathcal I)$ and a decreasing
sequence of non-negative eigenvalues
$\lambda_{1}\ge\lambda_{2}\ge \cdots \ge 0$ such that the covariance function in \ref{cov} can be represented as
\[
   \textrm{Cov}(x,y)=\sum_{p=1}^{\infty}\lambda_{p}\,\psi_{p}(x)\,\psi_{p}(y),
   \qquad x,y\in\mathcal I .
\]
From the Karhunen–Loève expansion \citep{karhunen1946spektraltheorie,loeve1946fonctions}, the functions $\hat{F}_1(x), \ldots, \hat{F}_N(x)$ can be expressed as a linear combination of the mean function $\bar{F}(x)$ and the set of orthonormal basis functions $\psi_p(x)$ multiplied by principal components \(\xi_{p,g}\), given by,
\[
   \xi_{p,g}= 
            \int_{\mathcal I} ( (\hat{F}_{g}(v)-\bar{F}(v)) -\,\psi_{p}(v) )\,\textrm{d}v
            = \langle \hat{F}_g-\bar{F},\psi_{p}\rangle_{L^{2}}.
\]
The scores are uncorrelated across $p$, with
$\E[\xi_{p,g}]=0$ and $\textrm{Var}(\xi_{p,g})=\lambda_p$. Specifically, the function $$\hat{F}_g(x) = \bar{F}(x) + \sum_{p=1}^{\infty}  \xi_{p,g} \psi_p(x),$$ for $g=1,\ldots,N,$ where $\psi_p(x)$ represents the $p^{th}$ dominant pattern of variation, $\xi_{p,g}$ denotes the amount of that pattern present in the particular curve $\hat{F}_g(x)$ and $\lambda_{p}$ is the variance explained by the $p^{th}$ pattern. In practice, the infinite sum is truncated at some $P$, retaining the first $P$ components that capture most of the variability. The proportion of variance explained by the $p$th component is estimated by
$
   \mathrm{PVE}(p)
      =\frac{\widehat{\lambda}_{p}}{ \sum_{j=1}^{P}\widehat{\lambda}_{j} }
       \times 100\%$, for
   $p=1,\dots,P$. 
 
Bivariate score plots, typically displaying either \(\bigl(\widehat{\xi}_{1g},\widehat{\xi}_{2g}\bigr)\) or \(\bigl(\widehat{\xi}_{1g},\widehat{\xi}_{3g}\bigr)\) for \(g=1,\ldots,N\) provide a useful visual tool for exploring data structure. Patterns such as clustering, elongated distributions, or isolated points may indicate latent groupings, underlying correlations, or potential outliers.

\subsubsection{Functional T-test}

Let \(F_{A_1},\dots,F_{A_{n_{A}}}\) and \(F_{B_1},\dots,F_{B_{n_{B}}}\) be independent samples of functions drawn from two populations with unknown mean functions \(\mu_A(x)\) and \(\mu_B(x)\), respectively. For each fixed \(x\in\mathcal I\subset\mathbb{R}\), we wish to test
\[
H_0:\ \mu_A(x)=\mu_B(x)\quad\text{versus}\quad H_1:\ \mu_A(x)\neq\mu_B(x).
\]
A pointwise two‑sample t‑statistic allowing for unequal variances is given by
\begin{equation}\label{T-test}
   T(x)
   =\frac{\hat\mu_A(x)-\hat\mu_B(x)}
          {\sqrt{\hat\sigma^2_A(x)/n_A+\hat\sigma^2_B(x)/n_B}},
\end{equation}
where \(\hat\mu_A(x)=\frac{1}{n_A}\sum_{i=1}^{n_A}F_{A_i}(x)\) and \(\hat\sigma^2_A(x)=\frac{1}{n_A-1}\sum_{i=1}^{n_A}\{F_{A_i}(x)-\hat\mu_A(x)\}^2\) (and analogously for group \(B\)) \citep{FDAB}. If equal variances can be assumed, one first computes the pooled variance estimator
\[
\hat S_p^2(x)=\frac{(n_A-1)\hat\sigma^2_A(x)+(n_B-1)\hat\sigma^2_B(x)}{n_A+n_B-2},
\]
and then replaces the denominator in \ref{T-test} by \(\sqrt{\hat S_p^2(x)\bigl(1/n_A+1/n_B\bigr)}\). To test the global null hypothesis \(H_0: \mu_A(x)=\mu_B(x)\) for all \(x\in\mathcal I\), we consider the supremum of the absolute value of the pointwise t‑statistic, $T_{\max}= \sup_{x\in\mathcal I} |T(x)|$. The null distribution of \(T_{\max}\) is approximated by a permutation procedure: randomly reassign the \(n_A+n_B\) functions to two groups of sizes \(n_A\) and \(n_B\), recompute \(T_{\max}\) for each permutation, and repeat many times. The observed \(T_{\max}\) is then compared to the resulting null distribution to obtain a \(p\)-value.

\subsubsection{Functional F-test}

Suppose we have $k$ independent groups of persistence landscapes, with $n_i$ functions in group $i$ denoted by $F_{i1}(x), F_{i2}(x), \ldots, F_{i n_i}(x)$ for $x\in\mathcal I\subset\mathbb{R}$, and total sample size $N = \sum_{i=1}^k n_i$. Let $\bar{F}_i(x) = \frac{1}{n_i}\sum_{j=1}^{n_i} F_{ij}(x)$ be the pointwise sample mean of the $i$th group, and let $\bar{F}(x) = \frac{1}{N}\sum_{i=1}^k\sum_{j=1}^{n_i} F_{ij}(x)$ be the overall pointwise mean.

We wish to test the null hypothesis that the group mean functions are equal everywhere,
\[
H_0:\; \bar{F}_1(x) = \bar{F}_2(x) = \cdots = \bar{F}_k(x) \quad \text{for all } x\in\mathcal I,
\]
against the alternative that at some $x$ at least two group means differ.

A pointwise F‑statistic, analogous to the classical one‑way ANOVA, is given by
\begin{equation}\label{eq:Ftest}
F(x) = \frac{\mathrm{SSR}(x)/(k-1)}{\mathrm{SSE}(x)/(N-k)},
\end{equation}
where the between‑group and within‑group sums of squares are
\[
\mathrm{SSR}(x) = \sum_{i=1}^k n_i \bigl(\bar{F}_i(x) - \bar{F}(x)\bigr)^2,
\qquad
\mathrm{SSE}(x) = \sum_{i=1}^k \sum_{j=1}^{n_i} \bigl(F_{ij}(x) - \bar{F}_i(x)\bigr)^2.
\]

To construct a global test, we use the supremum of the pointwise statistics, $F_{\max} = \sup_{x\in\mathcal I} F(x)$. Its null distribution is approximated by a permutation procedure: randomly reassign the $N$ functions to $k$ groups of sizes $n_1,\ldots,n_k$, recompute $F_{\max}$ for each permutation, and repeat many times. The observed $F_{\max}$ is then compared to this permutation distribution to obtain a $p$‑value.

\subsubsection{The funTDA framework}

The funTDA framework integrates persistent homology with functional data analysis (FDA) into a coherent pipeline for exploratory data analysis and statistical inference on network samples. As outlined in Algorithm~\ref{funTDA}, the procedure begins by computing for each network $G_g$ the geodesic distance matrix $\boldsymbol{\Gamma}_g$. Using this matrix, we examine how the network's structure evolves as a threshold parameter increases, recording when cycles (e.g., loops) appear and disappear. This information is summarized in a persistence diagram $\mathbf{P}_g$, which records the birth and death times of these topological features. Each diagram is then transformed into a persistence landscape $\{F_{g,(k)}(x)\}_{k =1}^{K},$ a collection of piecewise linear functions that reside in a Hilbert space and are amenable to FDA. Typically only the first layer $k=1$ is retained, as it captures the most dominant topological information.
The collection of landscapes $\{F_{g,(k)}\}_{g=1}^N$ is treated as a sample of functional objects. Standard FDA techniques including mean and covariance estimation, functional principal component analysis, and hypothesis testing are then applied to characterize the topological structure of the network sample and to draw inferential conclusions. These tools enable us to quantify typical topological patterns, identify dominant modes of variation across networks, and test for differences between groups. The complete workflow is summarized in Algorithm~\ref{funTDA}.

\begin{algorithm}
\caption{funTDA framework for statistical analysis of networks via persistent homology\label{funTDA}}
\begin{algorithmic}
\STATE \textbf{Input:} A sample of networks $\{G_g\}_{g=1}^N$ with adjacency matrices
$\{\mathbf{A}_g\}_{g=1}^N$.
\FOR{$g = 1,\ldots,N$}
    \STATE Compute the geodesic distance matrix
    $\boldsymbol{\Gamma}_g = (\gamma_{ij}^{(g)})$ associated with network $G_g$.
    \STATE Construct a filtration $\{G_{g,\epsilon}\}_{\epsilon}$ based on
    $\boldsymbol{\Gamma}_g$ and compute persistent homology.
    \STATE Extract the birth--death pairs corresponding to first-order homology,
    yielding the persistence diagram $\mathbf{P}_g$.
    \STATE Transform $\mathbf{P}_g$ into its persistence landscape representation
    $\{F_{g,(k)}(x)\}_{k = 1}^{K}$.
\ENDFOR
\STATE Treat $\{F_{g,(k)}\}_{g=1}^N$ as a sample of functions in a Hilbert space
and perform FDA, including estimation of mean and covariance
functions, functional principal component analysis, and hypothesis testing.
\STATE \textbf{Output:} Statistical summaries and inferential results characterizing
the topological structure of the networks.
\end{algorithmic}
\end{algorithm}

\section{Simulations}\label{sec3}

We conducted simulations of networks with diverse structures and sizes to evaluate the performance of our proposed method, which integrates persistent homology with functional data analysis for the statistical analysis of network samples. 

The first set of simulations generated samples of weighted, directed Erdős–Rényi random graphs, $G(n,p)$ \citep{erdos1959random}, where $n$ represents the number of vertices, and each directed edge exists independently with probability $p \in [0, 1]$ according to a Bernoulli distribution, $Ber(p)$. Edge weights were sampled independently from a $Beta(1,1)$ distribution. To examine the sensitivity of the method to connectivity, we considered three distinct network configurations defined by the edge probability: sparse networks with $p=0.1$, moderately connected networks with $p=0.5$, and dense networks with $p=0.9$. Example networks for each configuration, i.e., $Ber(0.1), Ber(0.5)$, and $Ber(0.9)$, are shown in Figure \ref{Fig7} in Appendix \ref{sim_ex}. For each configuration, we generated two cohort sizes: a smaller cohort of $N = 5$ networks and a larger cohort of $N = 20$. Section \ref{simpca}, examines how well bivariate score plots derived from our method could visually differentiate the six configurations. Section \ref{hyptest}, evaluates the ability of formal hypothesis tests to statistically identify differences among the six configurations.

In the second simulation, described in Section \ref{simcomparison}, we compared the proposed funTDA method against several existing approaches for network analysis. Specifically, we considered Quotient space methods and Laplacian-based techniques, including their Euclidean, Square Root, and Procrustes variants. The goal was to assess each method's ability to distinguish between two distinct network structures using two evaluation criteria: bivariate score plots for visual separation and t-tests for statistical group differences. Because Laplacian-based methods cannot be applied to directed graphs, we employed undirected, weighted Erdős–Rényi networks for this comparative analysis. This ensured that all methods could be evaluated on a common and appropriate network type.

In the third simulation in Section \ref{sbm}, we generated fully connected weighted undirected networks using a stochastic block model. This simulation was designed to emulate the real networks analysed later in the paper, namely the literary networks examined in Section \ref{literarycomparison} and the gene regulatory networks analysed in Section \ref{sec4}. In both of these applications the networks are undirected, weighted, and have equivalent density across networks within each sample, with variation arising from differences in interactions rather than the presence or absence of edges. Performance was evaluated based on the ability of funTDA to distinguish the network structures, assessed through visual separation in bivariate score plots and statistical testing of group differences using t-tests.

\subsection{Functional Principal Components Analysis}\label{simpca}

\begin{figure}[h!]
 \centering
 {{\includegraphics[width=14cm]{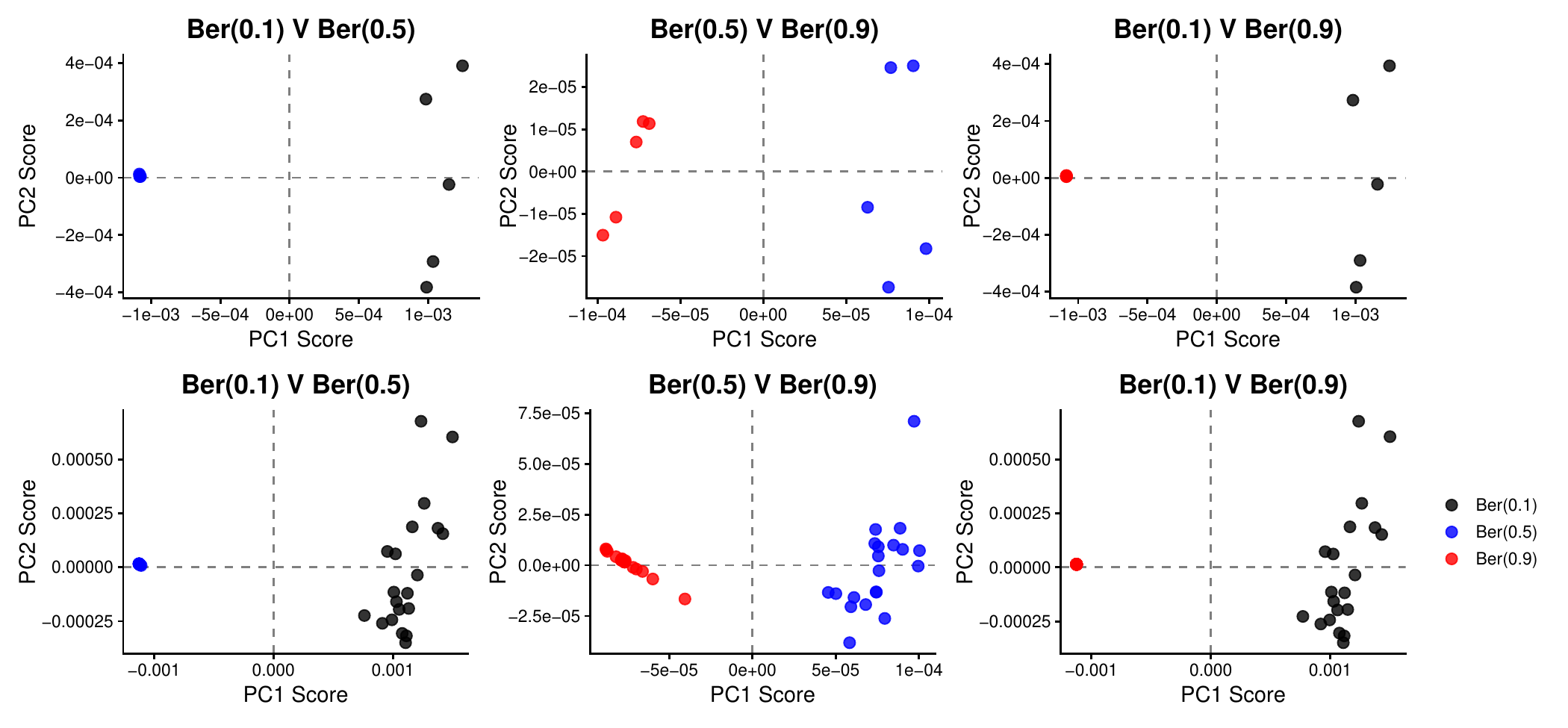} }}%
 \caption{Networks consisting of 1,000 vertices, with the top row representing a smaller sample size of N=5 and the bottom row corresponding to a larger sample size of N=20 networks of each type. \label{Fig12}}
 \end{figure}

Functional principal component analysis (FPCA) facilitates the identification of primary modes of variation within a sample of curves. In particular, bivariate score plots derived from FPCA can reveal potential groupings among observations. This simulation evaluates whether functional principal component (FPC) score plots can differentiate between networks generated from distinct connectivity structures. We generated networks with 1,000 vertices resulting in six distinct comparisons as described in Table \ref{tab:simcomparisons}.
\begin{table}[h]
\centering
\begin{tabular}{cl}
\hline
\textbf{Sample Size} & \textbf{Comparison} \\
\hline
\multirow{3}{*}{Small (\(N = 5\) per group)} 
    & (i) \(\text{Ber}(0.1)\) vs. \(\text{Ber}(0.5)\) \\
    & (ii) \(\text{Ber}(0.1)\) vs. \(\text{Ber}(0.9)\) \\
    & (iii) \(\text{Ber}(0.5)\) vs. \(\text{Ber}(0.9)\) \\
\hline
\multirow{3}{*}{Large (\(N = 20\) per group)} 
    & (iv) \(\text{Ber}(0.1)\) vs. \(\text{Ber}(0.5)\) \\
    & (v) \(\text{Ber}(0.1)\) vs. \(\text{Ber}(0.9)\) \\
    & (vi) \(\text{Ber}(0.5)\) vs. \(\text{Ber}(0.9)\) \\
\hline
\end{tabular}
\caption{Summary of simulation comparisons for small and large sample sizes.}
\label{tab:simcomparisons}
\end{table}
For each configuration in Table \ref{tab:simcomparisons}, we computed FPC scores and visualized them in bivariate score plots. Figure \ref{Fig12} presents these results, with the first principal component shown on the x-axis and the second on the y-axis. Networks are color-coded by their generating structure: black for \(\text{Ber}(0.1)\), blue for \(\text{Ber}(0.5)\), and red for \(\text{Ber}(0.9)\). The top row displays results for the smaller sample size (\(N = 5\)), while the bottom row corresponds to the larger sample size (\(N = 20\)). Across both sample sizes, networks cluster according to their underlying connectivity structure, as evidenced by the clear separation in FPC scores. These findings demonstrate that applying FPCA to topological summaries successfully distinguishes networks with connectivity structures.

\subsection{Hypothesis Testing}\label{hyptest}

To statistically assess differences among networks generated under varying connectivity structures, we conducted a series of hypothesis tests. Specifically, we compared networks each with $n=1,000$ and $n=250$ vertices sampled from three distinct connectivity structures: $Ber(0.1)$, $Ber(0.5)$, and $Ber(0.9)$. For each pair of structures, we applied a functional t-test to evaluate pairwise differences. We then employed a functional F-test to assess whether significant differences existed among all three connectivity structures simultaneously. Our expectation was that collections of networks drawn from the same distribution would exhibit no statistically significant differences, whereas those originating from different distributions would demonstrate clear separation.

Table \ref{tabhyptestsims} reports the average p-values from ten independent simulation runs, with standard deviations shown in parentheses. These p-values were derived from permutation tests using $10,000$ iterations and a significance level of $\alpha=0.05$. The simulations considered both smaller ($N=5$) and larger ($N=20$) samples of networks per structure.
 
 Consistent with our expectations, networks originating from different connectivity structures yielded average p-values below $0.05$, confirming statistically significant differences. Conversely, when testing networks drawn from the same distribution, we observed average p-values exceeding $0.05$, indicating no evidence of statistical differences. This pattern held across both sample sizes and vertex counts. Appendix \ref{simsizes2} presents additional analyses for scenarios in which the networks within each comparison had differing numbers of vertices, considering all possible combinations of 
$250$ and $1000$ vertices across the six simulation scenarios. These analyses produced comparable results and likewise identified statistically significant differences among the groups. Together, these findings demonstrate that the proposed approach successfully detects structural differences between the simulated networks.

\begin{table}[h!]
\caption{Presented are the mean p-values (standard deviations in parentheses) derived from executing ten simulations of a functional t-test for pairwise sample comparisons and a functional F-test for evaluating differences across three groups, each employing 10,000 permutations at a 0.05 significance threshold. If the value recorded is $<0.0001$ it is recorded as $0$.}\label{tabhyptestsims}
\begin{tabular*}{\textwidth}{@{\extracolsep\fill}lll}
\toprule%
Vertices = 1000 & 5 networks & 20 networks  \\
\midrule
Ber(0.1) V Ber(0.5)  & 0 (0)  & 0 (0)\\ 
Ber(0.1) V Ber(0.9)  &  0 (0)  & 0 (0) \\ 
Ber(0.5) V Ber(0.9)  &  0 (0) &  0 (0)\\ 
Ber(0.1) V Ber(0.5) V Ber(0.9) &   0 (0)  & 0 (0)\\
Ber(0.1) V Ber(0.1)  &  0.79 (0.18)   &  0.72 (0.18)\\
Ber(0.5) V Ber(0.5)  & 0.97 (0.07)  & 0.79 (0.15) \\
Ber(0.9) V Ber(0.9)  &  0.61 (0.42)  & 0.94 (0.11) \\
\midrule
Vertices = 250 \\
\midrule
Ber(0.1) V Ber(0.5)  &  0 (0)   & 0 (0)\\ 
Ber(0.1) V Ber(0.9)  &  0 (0) & 0 (0)\\ 
Ber(0.5) V Ber(0.9)  &  0 (0) &  0 (0)\\ 
Ber(0.1) V Ber(0.5) V Ber(0.9) &   0 (0)  & 0 (0)\\
Ber(0.1) V Ber(0.1)  & 0.76 (0.13)    & 0.66 (0.21)\\
Ber(0.5) V Ber(0.5)  &  0.56 (0.16)   & 0.82 (0.13) \\
Ber(0.9) V Ber(0.9)  &  0.78 (0.25)   & 0.47 (0.27)\\
\bottomrule
\end{tabular*}
\end{table}

\subsection{Comparison to Existing Methods on Simulated Networks }\label{simcomparison}

Two prominent methodologies for analyzing network samples, as discussed in the introduction, are the Quotient space method \citep{guo2021quotient} and the Laplacian method \citep{severn2022manifold}. In this simulation, we evaluate the proposed funTDA method against these approaches. For the Laplacian method, we consider its Euclidean, Square Root, and Procrustes variants. Our assessment focuses on each method's ability to distinguish between two network structures, using bivariate score plots for visual inspection and t-tests for statistical group comparisons. As the Laplacian approach cannot be applied to directed graphs, we conducted this comparative analysis using undirected, weighted Erdős–Rényi networks. We considered two connectivity patterns, $p=0.5$ and $p=0.9$, with networks comprising $250$ vertices. For each pattern, we generated both small ($N=5$) and large ($N=20$) samples to examine the influence of sample size on method performance.

\begin{figure}[h!]
 \centering
 {{\includegraphics[width=14cm]{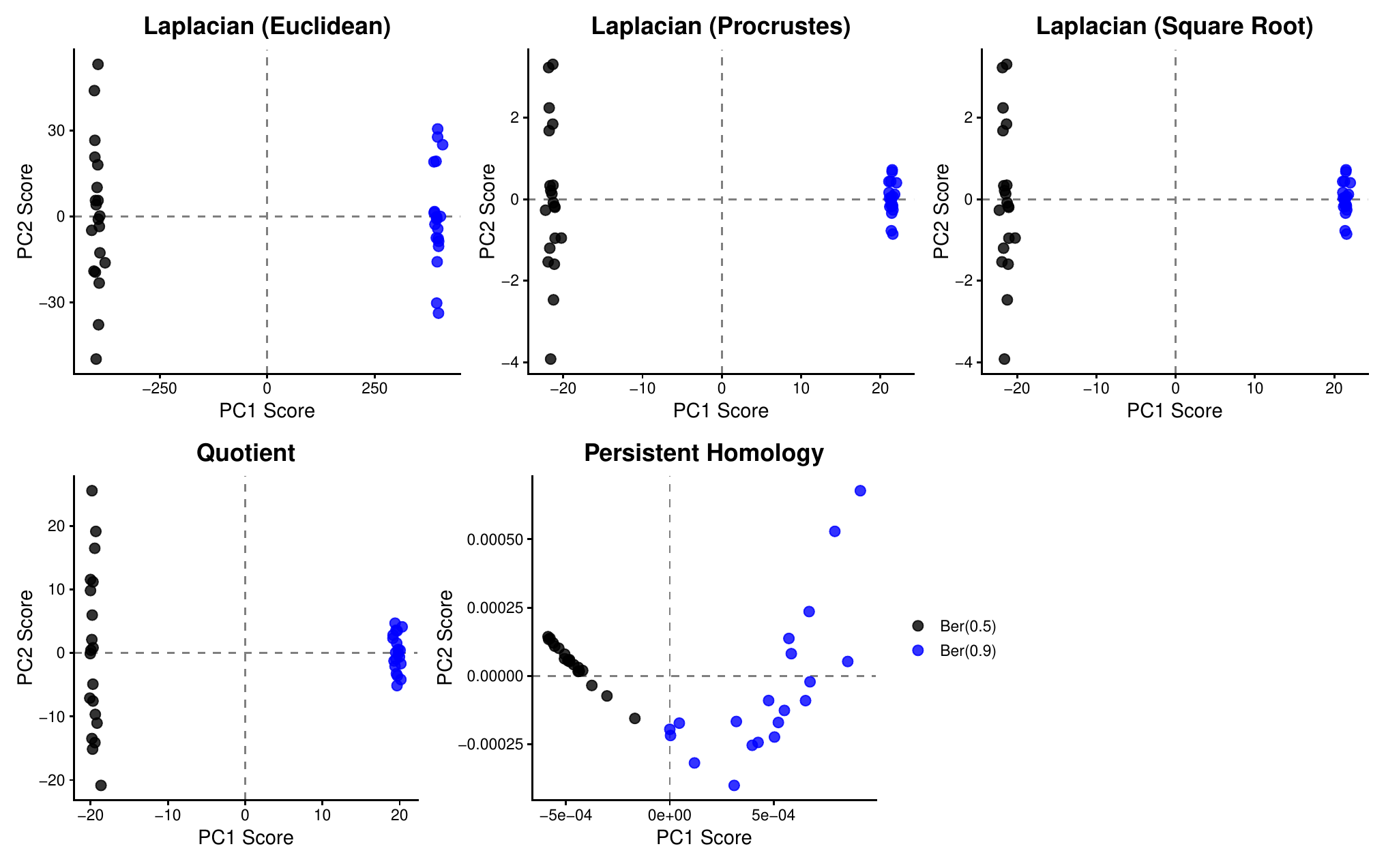} }}%
 \caption{Bivariate score plots with the x-axis showing the first principal component and the y-axis the second for the persistent homology, quotient and Laplacian approaches for networks consisting of $250$ vertices, with a sample size of $N=20$ networks of each type where Ber(0.5) is black, and Ber(0.9) is blue.\label{Fig14}}
 \end{figure}

Figure \ref{Fig14} presents the bivariate score plots for the larger sample ($N=20$), with the first principal component shown on the x-axis and the second on the y-axis. Networks generated from $Ber(0.5)$ are depicted in black, while those from $Ber(0.9)$ appear in blue. All methods achieved clear separation between the two groups in the principal component score space, demonstrating their capacity to distinguish networks with different connectivity structures. Appendix \ref{simsizes} shows comparable results with a clear separation between the two groups when $N=5$.

For hypothesis testing the proposed funTDA method and the Laplacian method (including its Euclidean, Square Root, and Procrustes variants) both permit t-test comparisons. However, the Quotient space method does not and is therefore excluded from this analysis. Additionally, because neither the Laplacian nor the quotient space method has an implementation for F-tests, these methods are also excluded from the F-test analysis.

The funTDA and Laplacian (Euclidean) methods were implemented using $10,000$ permutations at a significance level of $\alpha=0.05$.
However, the substantially higher computational cost associated with the Laplacian (Square Root) and Laplacian (Procrustes) variants rendered an equivalent number of permutations infeasible within a reasonable computational time-frame. Consequently, the number of permutations for these methods was reduced to $1000$ and $100$, refer to Appendix \ref{methodtiming} for computational timings of each approach. Computational times varied substantially across methods. funTDA was among the fastest methods, requiring 81 seconds (N=5) and 230 seconds (N=20). The Laplacian approaches ranged from 65 seconds (N=5) to 257 seconds (N=20) for the Euclidean variant, based on 10,000 permutations. In contrast, the Procrustes and square root variants were substantially more computationally expensive, with runtimes of 556 seconds (N=5) to 836 seconds (N=20), and 376 seconds (N=5) to 219 seconds (N=20), respectively, based on 100 permutations. Across both small ($N=5$) and larger ($N=20$) samples per network structure, the resulting permutation tests yielded p-values of zero to three decimal places, indicating that both methods consistently detected statistically significant differences between the two connectivity structures.

\subsection{Fully Connected Weighted Undirected Networks}\label{sbm}

To evaluate the ability of funTDA to detect differences in network structure independently of network density, we conducted a simulation study based on a weighted undirected stochastic block model. 
The design was chosen to emulate the real networks analysed later in the paper, namely the literary networks in Section~\ref{literarycomparison} and the gene regulatory networks in Section~\ref{sec4}, where the networks are undirected, weighted, and have comparable density across networks within each sample.

In this simulation, networks were generated with $n=250$ nodes and two alternative block configurations corresponding to $K\in\{2,5\}$ communities. Nodes were assigned to blocks of equal size. Specifically, when $K=2$ the networks consisted of two communities of 125 nodes each, whereas for $K=5$ the networks contained five communities of 50 nodes each. For each value of $K$, we generated both small ($N=5$) and larger ($N=20$) samples of networks. Networks were constructed as fully connected weighted undirected graphs. The block structure was introduced through the distribution of edge weights. Let $w_{ij}$ denote the weight of the edge between nodes $i$ and $j$. If nodes $i$ and $j$ belong to the same block, the edge weight is generated from a $\text{Beta}(6,1.5)$ distribution, whereas if the nodes belong to different blocks the weight is generated from a $\text{Beta}(1,4)$ distribution. These parameter choices produce larger weights within blocks than between blocks. In particular, the expected within-block weight is $\mathbb{E}(w_{ij}\mid \text{same block}) \approx0.8$,
while the expected between-block weight is $\mathbb{E}(w_{ij}\mid \text{different block})=0.2$. Consequently, all networks have identical density, since every edge is present, but differ in their underlying organisation through the block structures. This design allows us to assess whether funTDA can detect structural differences between networks independent of the network density.

\begin{figure}[t]
 \centering
 {{\includegraphics[width=12cm]{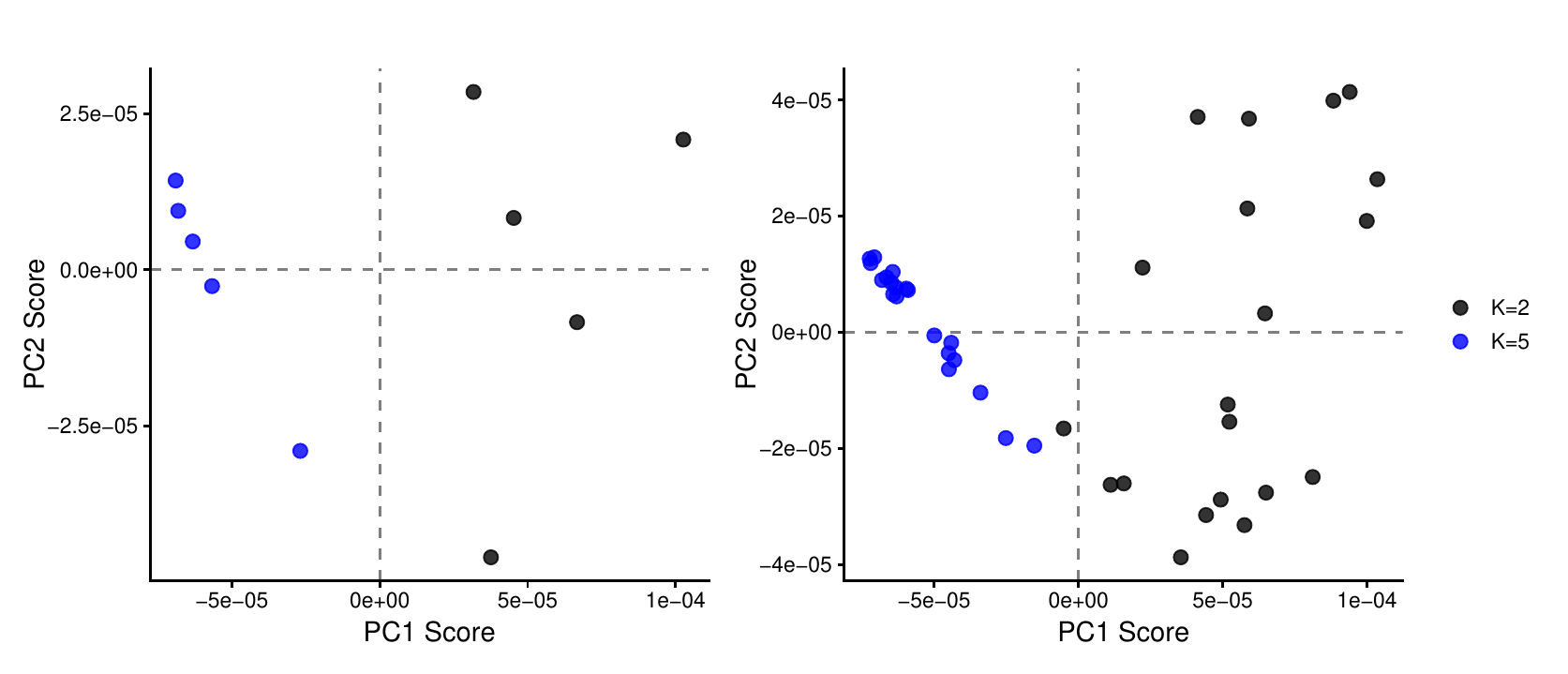} }}%
 \caption{Bivariate score plots for funTDA applied to stochastic block models with $K=2$ (black) and $K=5$ (blue), each consisting of 250 vertices. The x- and y-axes correspond to the first and second principal components, respectively.\label{FigSBM}}
 \end{figure}

Figure \ref{FigSBM} presents the bivariate score plots, with the first principal component shown on the x-axis and the second on the y-axis. The left panel displays the results for the smaller sample size $(N=5)$, while the right panel corresponds to the larger sample size $(N=20)$. Networks generated from $K=2$ are shown in black, whereas those generated from $K=5$ are shown in blue. Across both sample sizes, the networks cluster according to their underlying block structure, as indicated by the clear separation in the FPC scores. These results demonstrate that FPCA applied to the topological summaries is able to distinguish networks with different community structures.

Table \ref{tabhyptestsimssbm} summarises the average p-values obtained from ten independent simulation runs, with standard deviations reported in parentheses. The p-values were computed using permutation tests with $10,000$ iterations and a significance level of $\alpha = 0.05$. The simulations were conducted for both smaller ($N=5$) and larger ($N=20$) samples of networks per structure. As expected, comparisons between networks generated from different block configurations produced average p-values below $0.05$, indicating statistically significant differences. In contrast, when networks were sampled from the same block configuration, the resulting average p-values exceeded $0.05$, suggesting no evidence of structural differences. 

\begin{table}[h!]
\caption{Presented are the mean p-values (standard deviations in parentheses) derived from executing ten simulations of a functional t-test for pairwise sample comparisons, each employing 10,000 permutations at a 0.05 significance threshold. If the value recorded is $<0.0001$ it is recorded as $0$.}\label{tabhyptestsimssbm}
\begin{tabular*}{\textwidth}{@{\extracolsep\fill}lll}
\toprule%
Vertices = 250 & 5 networks & 20 networks  \\
\midrule
$K=2$ V $K=5$  & 0.00071 (0.0022)  & 0 (0)\\ 
$K=2$ V $K=2$  &  0.61 (0.17)  & 0.31 (0.24) \\ 
$K=5$ V $K=5$  &  0.49 (0.25) & 0.40 (0.22) \\ 
\bottomrule
\end{tabular*}
\end{table}

\section{Comparative Analysis of Literary Networks}\label{literarycomparison}

The literary network data introduced by \citet{severn2022manifold} consist of word co-occurrence networks constructed from the full texts of twenty-three novels, sixteen by Charles Dickens and seven by Jane Austen, obtained from the CLiC database \citep{mahlberg2016clic}. Each network encodes pairwise co-occurrence among the $n=250$ most frequent words across all novels. Nodes represent words, and weighted edges denote the number of times two words appear within a five-word window. Truncating to the $250$ most frequent words ensures a common vocabulary across all texts, yielding undirected networks with identical node sets across novels. This property is essential for methods such as the Laplacian approach, which require node correspondence across networks.
To account for differences in text length, each network's was normalized see Appendix \ref{norm} for details. These literary networks thus provide a natural and interpretable benchmark for comparing network-based methodologies, as their structures are thought to reflect distinctive linguistic and stylistic characteristics of the two authors.

\begin{figure}[h!]
 \centering
 {{\includegraphics[width=14cm]{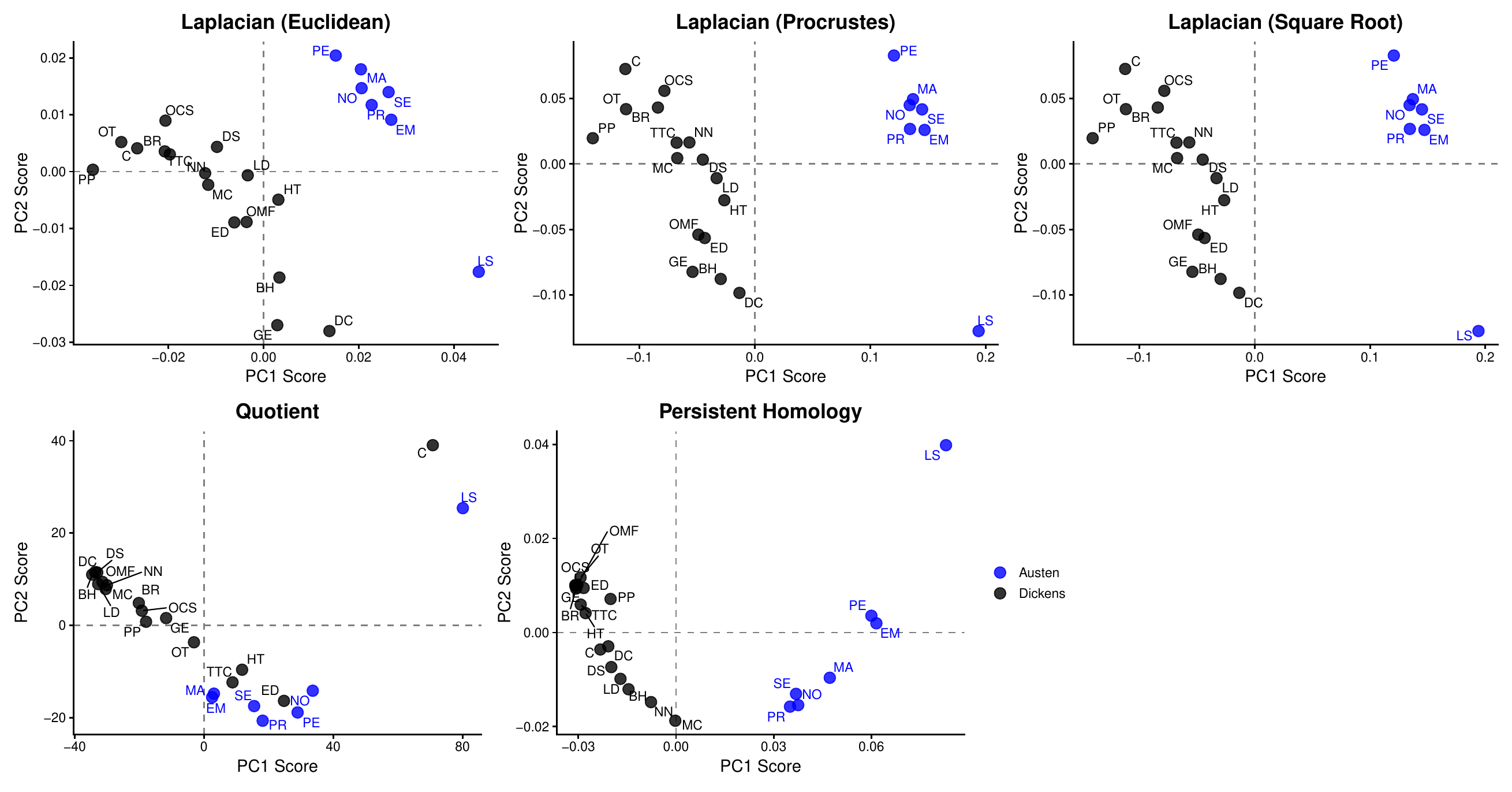} }}%
 \caption{Bivariate score plots for the persistent homology, quotient, and Laplacian approaches applied to literary networks of Dickens (black) and Austen (blue), each consisting of 250 vertices and a total of $N = 23$ networks. The x- and y-axes correspond to the first and second principal components, respectively.\label{Figliterary}}
 \end{figure}

We applied the proposed funTDA method, the Quotient space approach, and the Laplacian methods (Euclidean, Square Root, and Procrustes variants) to these networks. Figure~\ref{Figliterary} presents the resulting bivariate score plots, with the first principal component on the x-axis and the second on the y-axis. Blue dots represent Austen novels and black dots represent Dickens novels, with letter abbreviations corresponding to each novel, refer to Appendix \ref{novels}. Consistent with the findings of \citet{severn2022manifold}, both the persistent homology and Laplacian approaches yield clear separation between the two author groups. In contrast, the Quotient space method exhibits weaker separation between author groups, and its discrimination appears to be driven primarily by the distinctiveness of two particular novels: Dickens' A Christmas Carol (C) and Austen's Lady Susan (LS). These two works are well-documented as atypical within their respective authors' canons. Scholars have noted that A Christmas Carol deviates from Dickens's typical novelistic form and function \citep{moody2020implausible, zouidi2021sleight}, while Lady Susan is recognized as uncharacteristic of Austen's mature style and narrative voice \citep{russell2010hint}. This suggests that the Quotient space method may be detecting structural differences driven by these anomalous texts rather than capturing the broader stylistic distinctions between the two authors.

For both the funTDA persistent homology approach and all Laplacian variants, t-tests further confirm the differences between author groups, yielding p-values of zero to three decimal places. This indicates highly significant separation between the Dickens and Austen networks, consistent with the findings of \citet{severn2022manifold}. In contrast, the Quotient space method lacks a hypothesis testing framework and is therefore excluded from this test-based comparison.

\section{Gene Regulatory Network Differences Between Clinical Response Groups}\label{sec4}
In computational systems biology, a central goal is to model gene expression levels to understand how they respond to various external factors, and how these responses differ across patient groups.  Recent technological advancements now enable the monitoring of thousands of genes over time, and the Gene Expression Omnibus (GEO) \citep{edgar2002gene} provides a vast repository of such temporal expression data. We analyze data from \cite{huang2011temporal}, available on GEO under accession number GSE30550. In this study investigators examined gene expression in peripheral blood samples collected from 17 healthy participants following intranasal administration of H3N2 influenza virus. For each subject, expression profiles for 22,277 genes were measured at 15 time points: 0, 5, 12, 21.5, 29, 36, 45.5, 53, 60, 69.5, 77, 84, 93.5, 101, and 108 hours post-infection. All participants tested negative via rapid antigen detection (BinaxNow Rapid Influenza Antigen; Inverness Medical Innovations, Inc) at the final 108-hour time point. In addition to gene expression data, symptomatic scores and viral shedding measurements were recorded for each subject throughout the experimental period. Based on these clinical metrics, eight subjects were classified as asymptomatic and nine as symptomatic. Distinguishing between these two groups at the transcriptional level is essential for identifying the gene expression signatures associated with protective immunity, discovering early diagnostic biomarkers, and revealing potential therapeutic targets that promote viral clearance while minimizing immunopathology.

Following the preprocessing and differential expression pipeline outlined in \cite{carey2018big}, we identified for each subject the 250 genes exhibiting the most significant temporal expression changes. For each subject, we then constructed a gene regulatory network (GRN) by computing pairwise Spearman rank correlations among the temporal expression profiles of these 250 genes. This procedure yielded 17 correlation matrices, each of size $250 \times 250$, which serve as the weighted adjacency matrices for the respective GRNs. As the set of top 250 genes is subject-specific, the resulting networks are weighted and undirected but lack shared node correspondence across individuals.

\begin{figure}[h!]
 \centering
 {{\includegraphics[width=14cm]{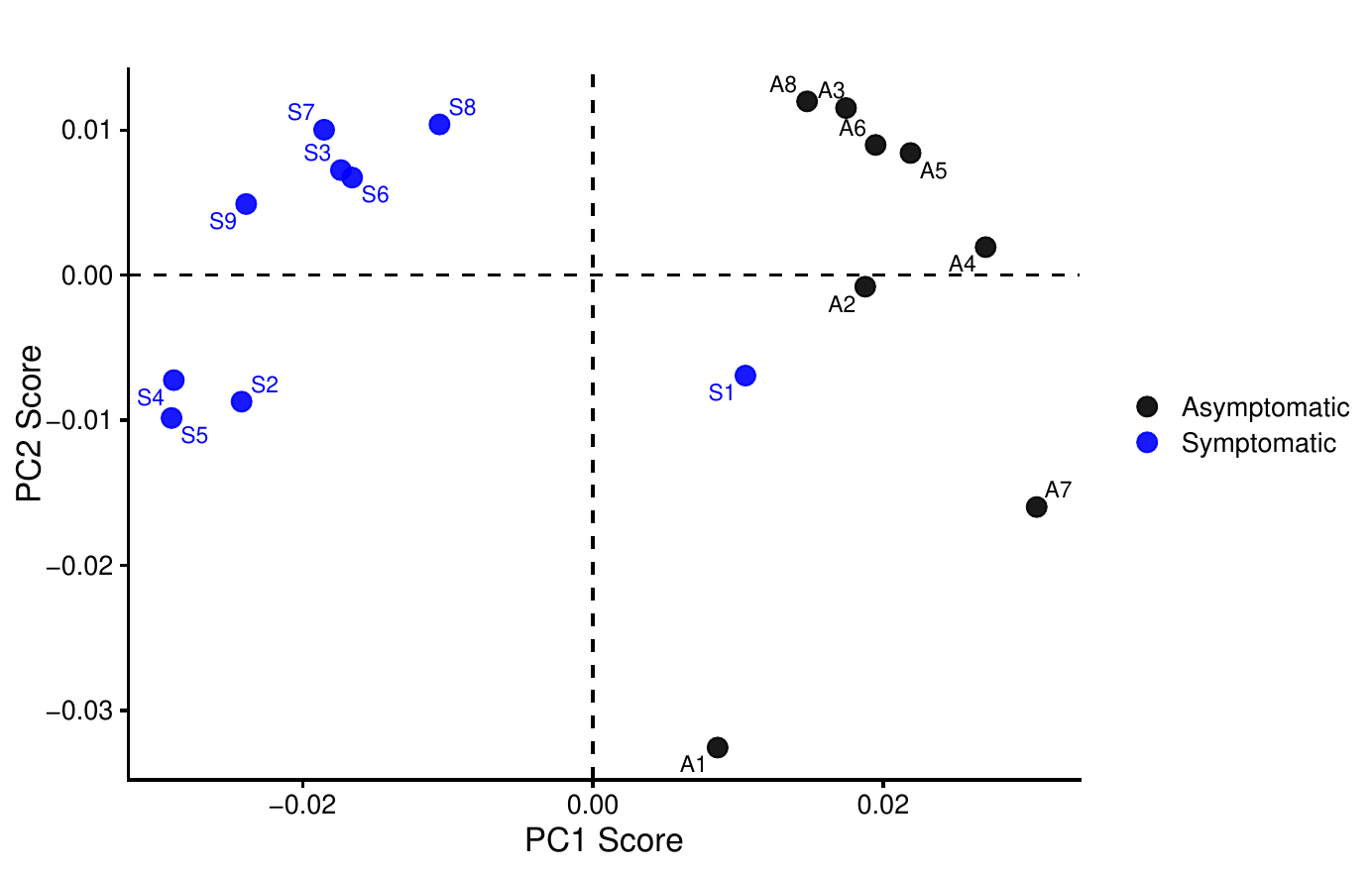} }}
 \caption{The scores for the first two principal components denoted by ‘A' for asymptomatic, and ‘S' for symptomatic subjects GRNs. \label{Fig19}}
 \end{figure}

We then performed FPCA on the collection of persistence landscapes derived from the combined set of symptomatic and asymptomatic networks. The results, illustrated in Figure \ref{Fig19}, reveal a separation between the two groups: networks from asymptomatic subjects, denoted by `A' (black), cluster separately from those of symptomatic subjects, denoted by `S' (blue). This separation indicates that the topological structures of gene regulatory networks differ between asymptomatic and symptomatic individuals. The finding suggests that the proposed funTDA approach can effectively capture biologically meaningful distinctions between the two groups. In all analyses, we examine the features with the highest persistence level, $k=1$. In Appendix \ref{klandscapes}, we further examine the FPC scores for progressively lower persistence levels, 
$k=2,3,4,$ and $5$, representing a decreasing order of prominence. As shown in Figure \ref{klands}, in Appendix \ref{klandscapes} the clear separation between the two groups is maintained across these lower persistence levels.

To formally assess whether a significant difference exists between the networks of symptomatic and asymptomatic subjects, we conducted a permutation hypothesis test with 10,000 permutations at a significance level of $\alpha=0.05$. As detailed in Table \ref{Tab3}, the analysis of persistence landscapes yielded a p-value of $0.001$. This result indicates a statistically significant difference between the two groups, suggesting that the topological properties of the gene regulatory networks vary according to clinical response. We further examined the hypothesis tests at progressively lower persistence levels, 
$k=2,3,4,$ and $5$, which represent decreasing levels of prominence. All tests remained significant at the $\alpha=0.05$ level, providing consistent evidence of differences between the two groups.

As a control analysis, we also investigated whether networks within the same clinical response group exhibit significant differences. Within each subgroup (asymptomatic and symptomatic), the networks were divided into two halves, and the hypothesis test was performed for every possible such partition. As expected, the mean p-values exceeded $0.05$, indicating no significant differences among networks from subjects sharing the same clinical response. 

\begin{table}[h!]
\caption{Permutation test with 10,000 permutations at 0.05 level for asymptomatic and symptomatic GRNs,  (standard deviations in parentheses).}\label{Tab3}
\small
\begin{tabular*}{\textwidth}{@{\extracolsep\fill}ll}
\toprule%
 &  p-value \\
\midrule
Symptomatic V Asymptomatic &   0.001 \\
Symptomatic V Symptomatic &   0.495 (0.2896) \\ 
Asymptomatic V Asymptomatic & 0.4792 (0.2904) \\ 
\bottomrule
\end{tabular*}
\end{table}

\section{Conclusion}\label{sec5}

This paper introduces funTDA, a novel approach for the statistical analysis of collections of networks that integrates topological data analysis with functional data analysis. We employ persistent homology to extract topological features from networks, focusing specifically on first-order properties, commonly interpreted as "loops", and represent these features using persistence landscapes. These functional representations enable the application of standard FDA techniques, such as principal component analysis and hypothesis testing, to network data. 

Our approach circumvents several challenges inherent to existing methods. It avoids the difficulties associated with covariance estimation and inversion, and overcomes the computational costs of graph matching procedures. Moreover, the proposed method does not require networks to have equal sizes or known node correspondences, and it accommodates a wide range of network types, including weighted and unweighted, as well as directed and undirected networks. Notably, among methods that do not require node correspondence, funTDA uniquely enables formal hypothesis testing.

We validated the performance of our approach through an extensive simulation study. In Erdős–Rényi simulations, funTDA successfully distinguished between different network structures, as evidenced by both FPCA visualizations and hypothesis tests. Compared to the Quotient space method, persistent homology achieved comparable separation in principal component scores; however, the Quotient approach lacks a hypothesis testing framework and is constrained by computationally demanding graph matching, which limits the number of vertices and networks that can be analyzed. Relative to the Laplacian approach, persistent homology yielded comparable separation in principal component scores and statistically significant p-values. In the stochastic block model simulations, funTDA effectively distinguished networks with different underlying structures while remaining insensitive to overall network density, as demonstrated by the clear separation in principal component scores and statistically significant p-values. In the literary network analysis, funTDA clearly separated Dickens and Austen networks in bivariate score plots and produced highly significant p-values, performing comparably to the Laplacian approach.

The proposed method is particularly valuable in fields such as computational biology, where detecting differences between gene regulatory networks can yield insights into underlying biological processes and disease mechanisms. In our application to H3N2 influenza GRNs, the approach successfully distinguished between symptomatic and asymptomatic subjects, revealing clear separation in principal component scores and identifying statistically significant differences between groups, while detecting no significant differences among networks from subjects within the same clinical response category.

Beyond these specific applications to literary and gene regulatory networks, the proposed approach provides a general framework for analysing complex networks in which node correspondence is not defined, a common situation in settings such as social networks and transportation networks. Many existing approaches, including Laplacian-based methods, rely on direct node-to-node comparisons and are therefore not applicable when networks differ in size or composition. Moreover, alternative approaches that do not require node correspondence, such as the quotient space method, lack a formal hypothesis testing framework and is computationally demanding, as it requires storing all possible permutations in the quotient space. By combining topological invariants with functional representations, the proposed funTDA persistent homology–based approach addresses these limitations and highlights the potential of topological data analysis for advancing the statistical analysis and understanding of complex network-based datasets. An implementation of the method is freely available at \url{https://github.com/CatherineH1/funTDA}.

Finally, several directions for future research merit further investigation. In particular, extending the funTDA framework to incorporate clustering methodologies \citep{jacques2014functional, higgins2024addressing} could enable the identification of distinct network subtypes within a sample. Similarly, integrating supervised classification approaches \citep{wang2024review} may facilitate the prediction of group membership, such as disease versus control status, based on network structure. However, developing these extensions would require substantial additional methodological work, and therefore lies beyond the scope of the present study.

\begin{acks}[Acknowledgments]

\end{acks}

\begin{funding}
This research was funded by Research Ireland through the Research Ireland Centre for Research Training in Genomics Data Science under grant number 18/CRT/6214.
\end{funding}

\begin{supplement}
\stitle{Code and Data}
\sdescription{The code and data are
available at \url{https://github.com/CatherineH1/funTDA}}
\end{supplement}

\newpage
\appendix

\section{$H_0$ features}\label{H_0}

In the example shown in Figure~\ref{Fig1}, all
vertices are born at filtration value \(0\), initially yielding five connected components. As the filtration progresses, components merge: at \(\epsilon=0.1\) vertices A and B connect (four components remain); at \(\epsilon=0.3\) vertices D and E connect (three components); at
\(\epsilon=0.4\) vertices B and C connect (two components); and at \(\epsilon=0.5\) the remaining components merge into a single connected component. This evolution is represented in the persistence diagram by the red points, whose birth times lie at \(0\) and whose death times indicate the filtration values at which component merges occur.

\section{An illustrative example of the Erdős–Rényi random graphs}\label{sim_ex}

\begin{figure}[h!]
 \centering
 {{\includegraphics[scale=0.18]{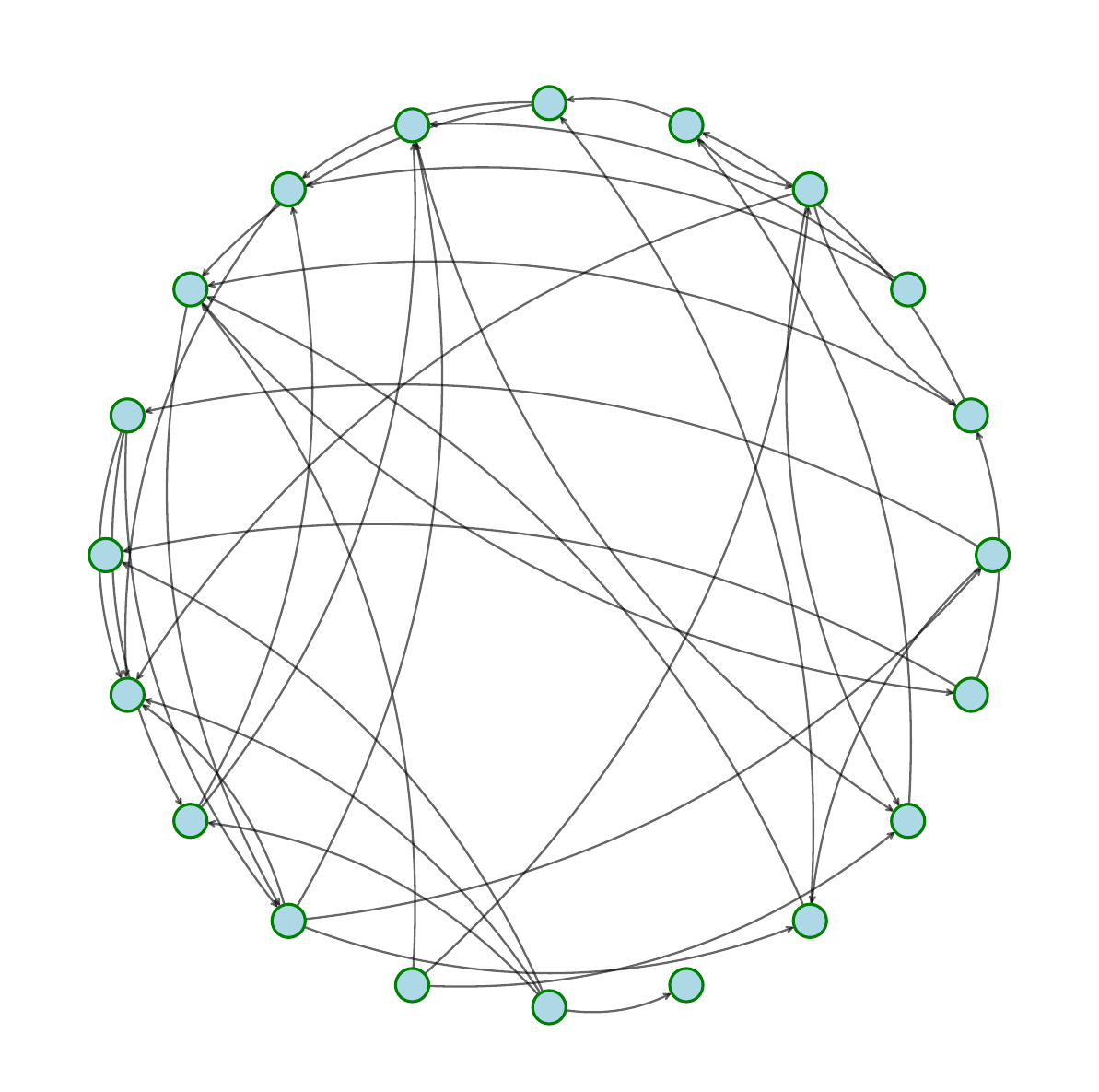} }}%
 \qquad
 {{\includegraphics[scale=0.18]{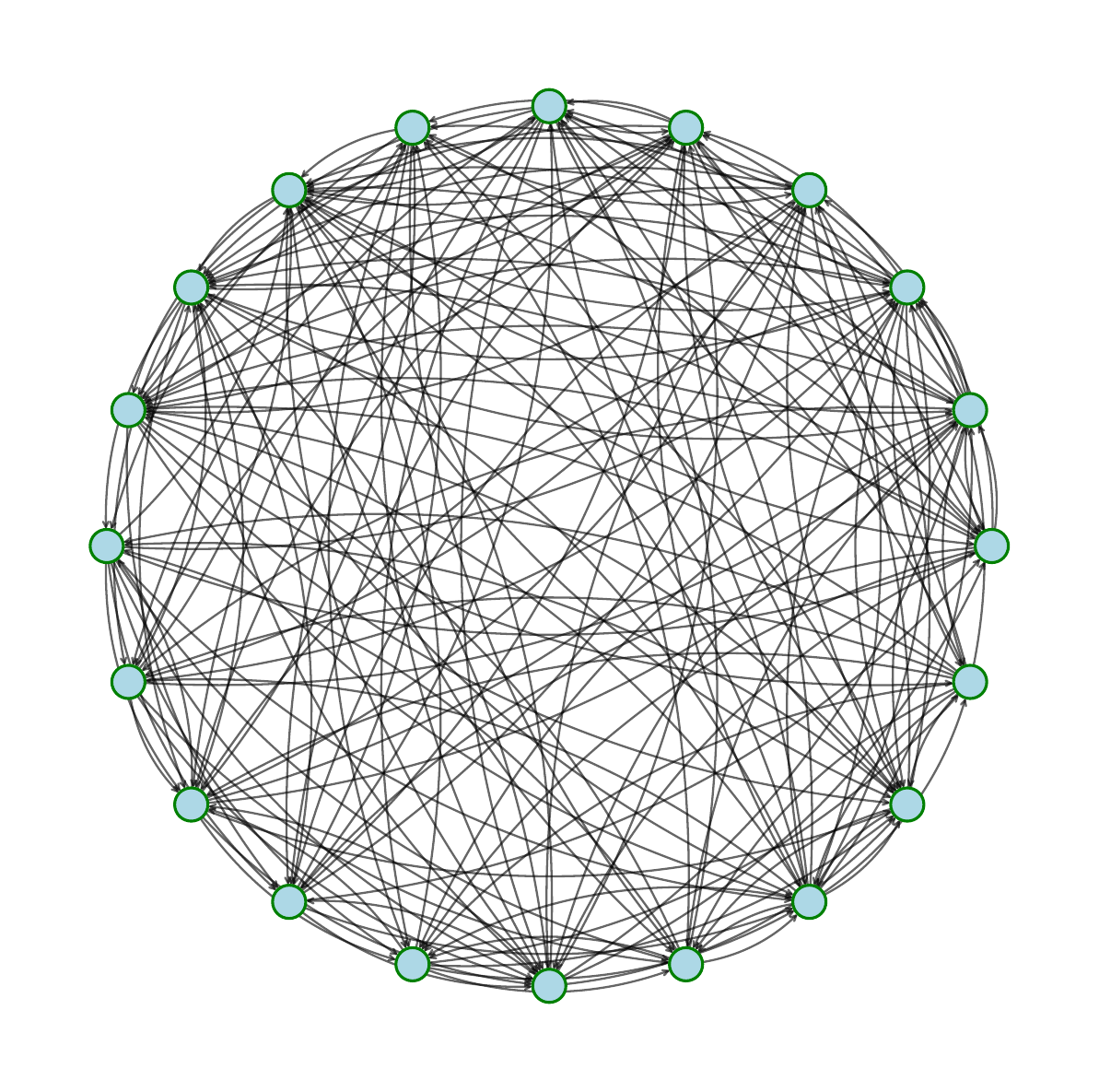} }}%
 \qquad
 {{\includegraphics[scale=0.18]{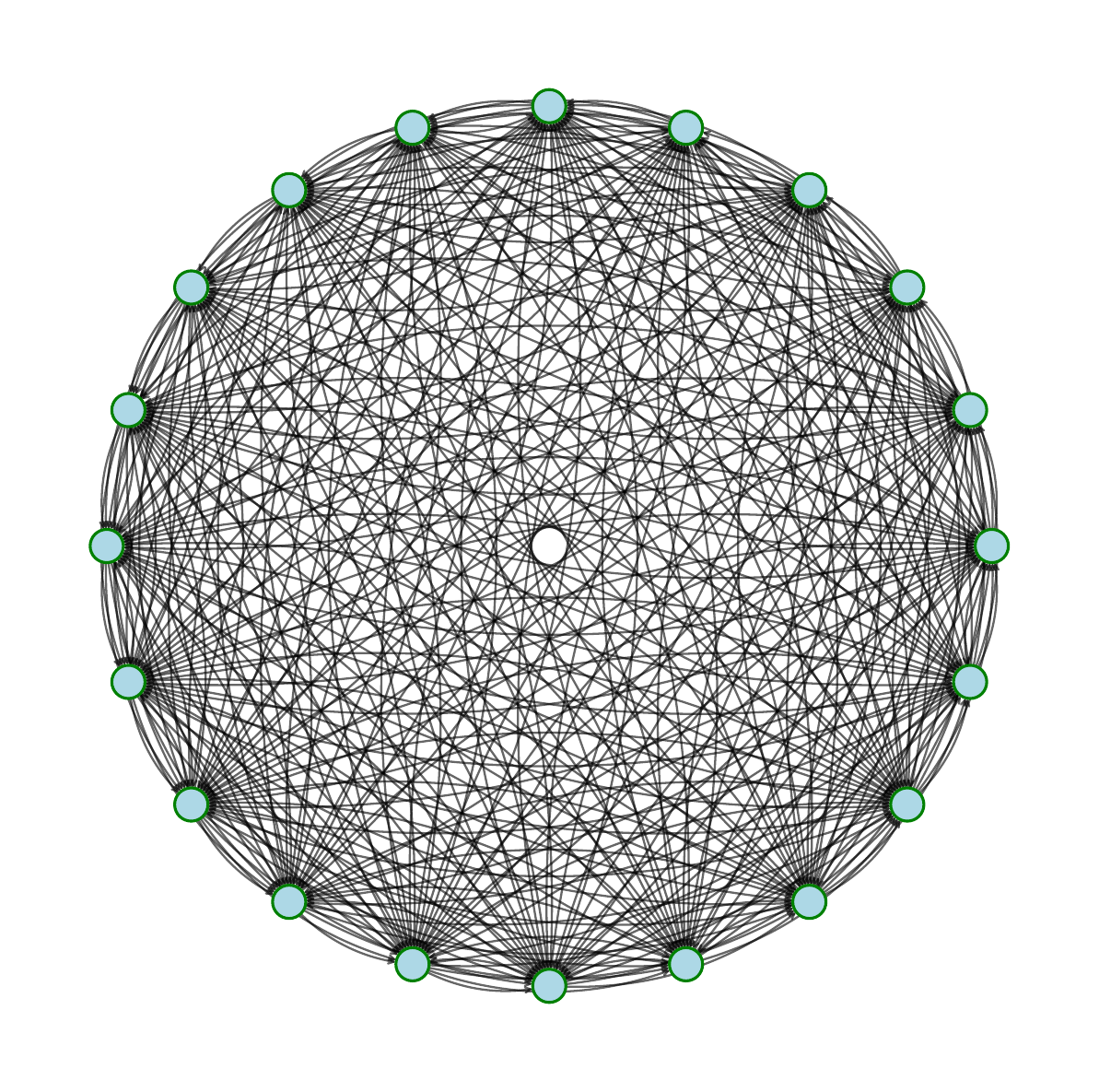} }}%
 \caption{An illustrative example of a network from each connectivity structure type, $Ber(0.1)$ left, $Ber(0.5)$ middle, and $Ber(0.9)$ right.  \label{Fig7}}
 \end{figure}

The first set of simulations generated samples of weighted, directed Erdős–Rényi random graphs, $G(n,p)$ \citep{erdos1959random}, where $n$ represents the number of vertices, and each directed edge exists independently with probability $p \in [0, 1]$ according to a Bernoulli distribution, $Ber(p)$. Edge weights were sampled independently from a $Beta(1,1)$ distribution. To examine the sensitivity of the method to connectivity, we considered three distinct network configurations defined by the edge probability: sparse networks with $p=0.1$, moderately connected networks with $p=0.5$, and dense networks with $p=0.9$. Example networks for each configuration, i.e., $Ber(0.1), Ber(0.5)$, and $Ber(0.9)$, are shown in Figure \ref{Fig7} with $Ber(0.1)$ left, $Ber(0.5)$ middle, and $Ber(0.9)$ right. 

\section{Simulations with Varying Numbers of Vertices Between Samples}\label{simsizes2}
In this simulation, we consider networks with varying numbers of vertices between samples. The networks were generated with either 250 or 100 vertices for all possible combinations across the six simulation scenarios.  We generate collections of directed, weighted Erdős–Rényi random graphs, $G(n,p)$, as described in Section \ref{sec3}. We consider two sample sizes: a small sample of $N = 5$ networks and a larger sample of $N = 20$ networks of each type.

Table \ref{tabdiffnodes} presents the average p-values from ten independent simulation runs, with standard deviations shown in parentheses. These p-values were derived from permutation tests using $10,000$ iterations and a significance level of $\alpha=0.05$. The simulations considered both smaller ($N=5$) and larger ($N=20$) samples of networks per structure. The low p-values indicate a statistically significant difference between networks of different structures, where the networks have varying numbers of vertices. The results show that the proposed approach successfully differentiates networks of different structures, even when they have different vertex counts between samples.

\begin{table}[h!]
\caption{Presented are the mean p-values (standard deviations in parentheses) derived from executing ten simulations of a functional t-test for pairwise sample comparisons and a functional F-test for evaluating differences across three groups, each employing 10,000 permutations at a 0.05 significance threshold for directed weighted networks with different numbers of vertices between samples. If the value recorded is $<0.0001$ it is recorded as $0$.}\label{tabdiffnodes}
\footnotesize
\begin{tabular*}{\textwidth}{@{\extracolsep\fill}lcc}
\toprule%
 & 5 networks & 20 networks\\
\midrule

Ber(0.1) n=250 V Ber(0.5) n=1000  &  0 (0) & 0 (0)  \\ 
Ber(0.1) n=1000 V Ber(0.5) n=250  & 0 (0) & 0 (0) \\ 
Ber(0.1) n=1000 V Ber(0.9) n=250  & 0 (0) & 0 (0) \\ 
Ber(0.1) n=250 V Ber(0.9) n=1000  & 0 (0) & 0 (0) \\ 
Ber(0.5) n=1000 V Ber(0.9) n=250  & 0 (0) & 0 (0) \\ 
Ber(0.5) n=250 V Ber(0.9) n=1000  & 0 (0) & 0 (0) \\ 
Ber(0.1) n=250 V Ber(0.5) n=1000 V Ber(0.9) n=1000 & 0 (0)  &  0 (0)\\ 
Ber(0.1) n=250 V Ber(0.5) n=250 V Ber(0.9) n=1000 & 0 (0)  &  0 (0)\\ 
Ber(0.1) n=250 V Ber(0.5) n=1000 V Ber(0.9) n=250 & 0 (0)  &  0 (0)\\ 
Ber(0.1) n=1000 V Ber(0.5) n=250 V Ber(0.9) n=1000 & 0 (0)  &  0 (0)\\ 
Ber(0.1) n=1000 V Ber(0.5) n=1000 V Ber(0.9) n=250 & 0 (0)  &  0 (0)\\ 
Ber(0.1) n=1000 V Ber(0.5) n=250 V Ber(0.9) n=250 & 0 (0)  &  0 (0)\\

\bottomrule
\end{tabular*}
\end{table}

\section{Method comparison when N=5}\label{simsizes}
\begin{figure}[h!]
 \centering
 {{\includegraphics[width=14cm]{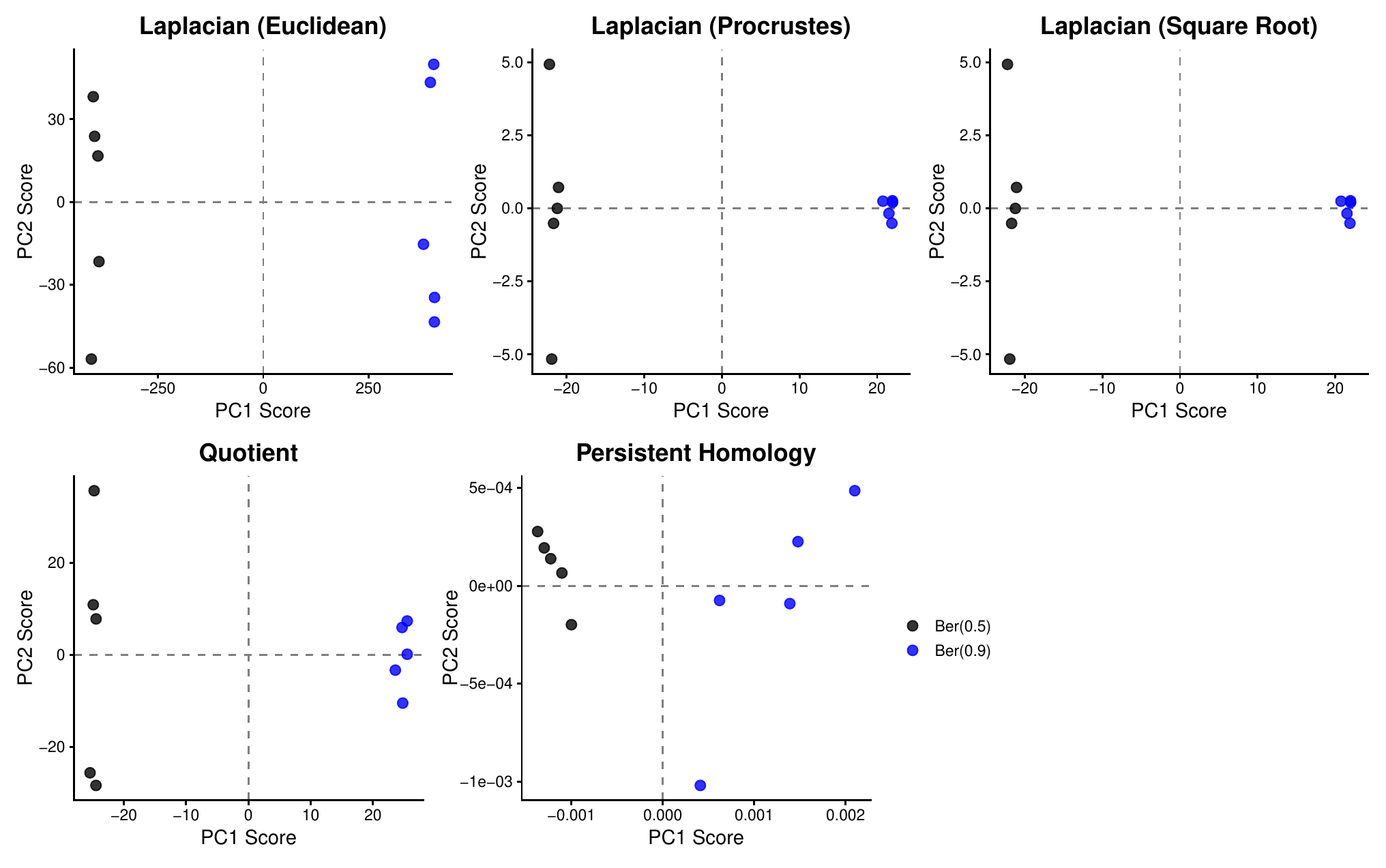} }}%
 \caption{Bivariate score plots with the x-axis showing the first principal component and the y-axis the second for the persistent homology, quotient and Laplacian approaches for networks consisting of $250$ vertices, with a sample size of $N=5$ networks of each type where Ber(0.5) is black, and Ber(0.9) is blue.\label{comparisonnis5}}
 \end{figure}

This simulation examines networks with two different connectivity patterns, $Ber(0.5)$ and $Ber(0.9)$, for a small ($N = 5$) sample, each consisting of $250$ vertices. Figure \ref{comparisonnis5} presents the bivariate score plots, with the x-axis showing the first principal component and the y-axis the second for the small sample (\( N = 5 \)). Black dots depict $Ber(0.5)$ while blue dots indicate $Ber(0.9)$. All approaches achieve clear separation in principal component scores between the two groups.

\section{Normalization to account for text length}\label{norm}

To account for differences in text length, each network's graph Laplacian was normalized by its trace in the Laplacian method, producing a unit-trace representation that removes the influence of text length. For the proposed funTDA persistent homology approach and the Quotient space method, adjacency matrices were normalized by their maximum entry, rescaling all edge weights to the interval $[0,1]$. This normalization preserves the relative connection structure within each network while ensuring comparability in scale, analogous to the trace normalization employed by \cite{severn2022manifold} for the Laplacian method.

\section{Additional Persistence Landscape Levels}\label{klandscapes}
In the main analysis we focus on the first persistence landscape level ($k=1$), which corresponds to the largest landscape function and therefore captures the most persistent and globally dominant topological features of the persistence diagram. This level represents the pointwise maximum across all triangular functions constructed from the birth-death pairs and is commonly used in practice as it summarizes the most prominent global topological structures.

\begin{figure}[h!]
 \centering
 {{\includegraphics[width=11cm]{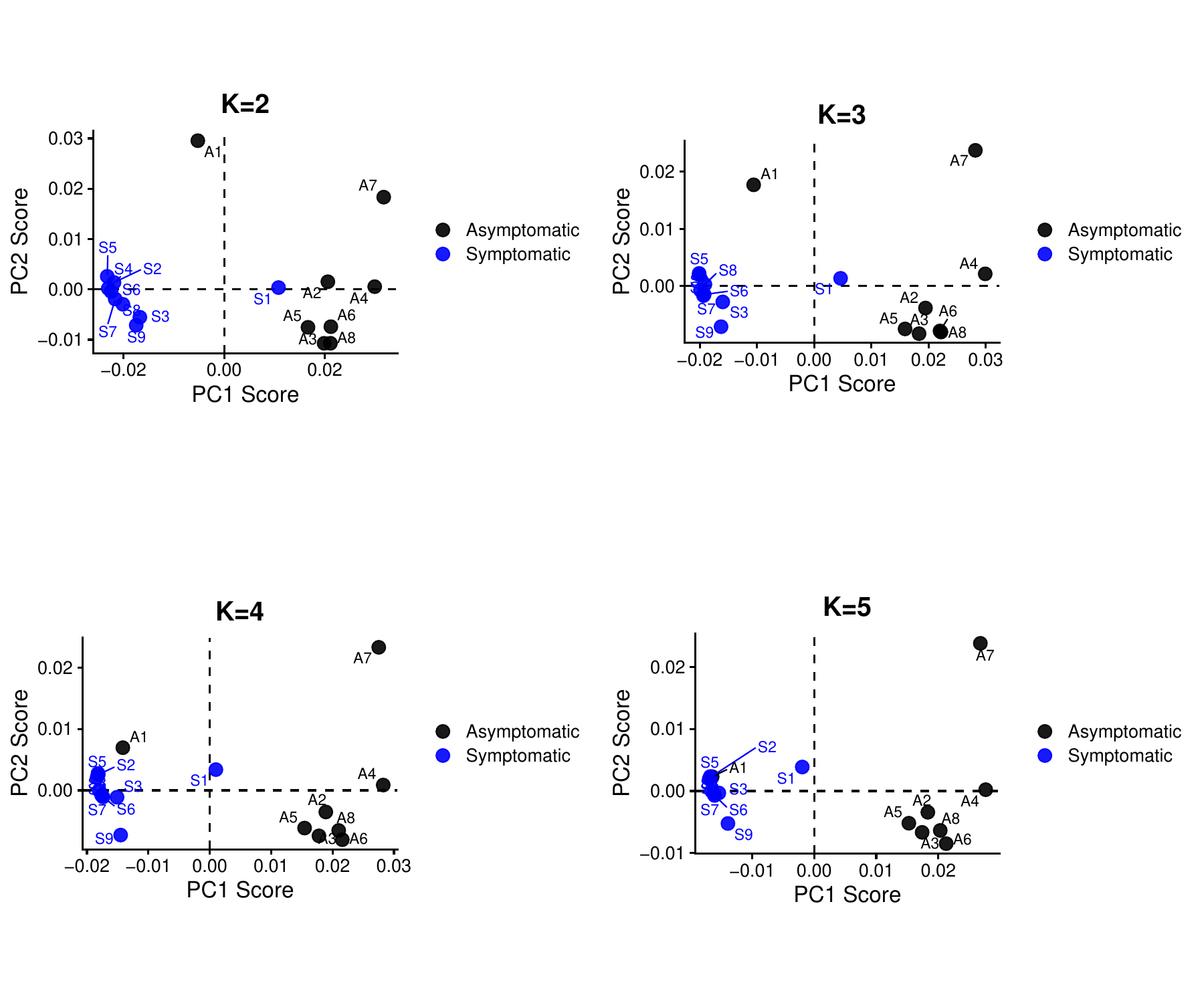} }}%
 \caption{Bivariate principal component score plots derived from persistence landscapes using levels $k=2,\ldots,5$ for the gene regulatory networks. Across all landscape levels there is a clear separation between the two groups.\label{klands}}
 \end{figure}

To assess whether the results depend on this choice, we also examined additional landscape levels ($k=2,3,4,5$). These levels represent progressively less persistent topological features. In the gene regulatory network analysis presented in Section~\ref{sec4}, incorporating these additional landscape levels produced results that were similar to those obtained for 
$k=1$. Specifically, a clear separation in the principal component scores between the two groups was observed across all $k$ levels considered (see Figure~\ref{klands}).

\section{Method Execution Times}\label{methodtiming}
Table~\ref{tabdiffnodes} reports the computational time in secods required to perform the principal component analysis and subsequent hypothesis testing for each method. Timings are presented for both smaller ($N=5$) and larger ($N=20$) samples of networks. For hypothesis testing, the reported times reflect the number of permutations required by each method. 

\begin{table}[h!]
\caption{Computation times (in seconds) for the permutation-based hypothesis testing across the different network analysis methods. Results are reported for samples of $N=5$ and $N=20$ networks. The number of permutations used for each hypothesis test is indicated in parentheses}\label{tabdiffnodes}
\footnotesize
\begin{tabular*}{\textwidth}{@{\extracolsep\fill}lcc}
\toprule%

t-test \\
\midrule
funTDA &  81.10 (10000 permutations) & 230.14 (10000 permutations)\\
Laplacian (Euclidean) & 65.97 (10000 permutations)& 257.16 (10000 permutations)\\
Laplacian (Procrustes) &556.39 (100 permutations)  & 836.91 (100 permutations) \\
Laplacian (Square Root) &376.99(1000 permutations) & 219.81 (100 permutations)\\

\bottomrule
\end{tabular*}
\end{table}

\section{CLiC Novel Details}\label{novels}
This appendix lists the twenty-three novels analyzed, including their full titles and corresponding abbreviations, comprising sixteen by Charles Dickens and seven by Jane Austen, obtained from the CLiC corpus \citep{mahlberg2016clic}.

\begin{table}[h!]
\caption{The Jane Austen and Charles Dickens novels from the CLiC database \citep{mahlberg2016clic}}
\centering
\begin{tabular*}{\textwidth}{@{\extracolsep\fill}lcc}
\hline
\textbf{Author} & \textbf{Novel name} & \textbf{Abbreviation} \\
\hline 
Austen & Lady Susan & LS\\
Austen & Sense and Sensibility & SE\\
Austen & Pride and Prejudice & PR\\
Austen & Northanger Abbey & NO\\
Austen & Mansfield Park & MA\\
Austen & Emma & EM\\
Austen & Persuasion & PE\\
Dickens & The Pickwick Papers & PP\\
Dickens & Oliver Twist & OT\\
Dickens & Nicholas Nickleby & NN\\
Dickens & The Old Curiosity Shop & OCS\\
Dickens & Barnaby Rudge & BR\\
Dickens & Martin Chuzzlewit & MC\\
Dickens & A Christmas Carol & C\\
Dickens & Dombey and Son & DS\\
Dickens & David Copperfield & DC\\
Dickens & Bleak House & BH\\
Dickens & Hard Times & HT\\
Dickens & Little Dorrit & LD\\
Dickens & A Tale of Two Cities & TTC\\
Dickens & Great Expectations & GE\\
Dickens & Our Mutual Friend & OMF\\
Dickens & The Mystery of Edwin Drood & ED\\
\bottomrule
\end{tabular*}
\end{table}

\bibliographystyle{imsart-nameyear} 
\bibliography{bibliography.bib}       

@book{FDAB,
  author = {Ramsay, J. O. and Silverman, B. W.},
  publisher = {Springer},
  title = {Functional Data Analysis},
  year = 2005
}

@article{moody2020implausible,
  title={" Implausible and Inappropriate"? A defense of Dickens's A Christmas Carol},
  author={Moody, Emily},
  journal={Dickens Quarterly},
  volume={37},
  number={4},
  pages={393--408},
  year={2020},
  publisher={Johns Hopkins University Press}
}

@article{russell2010hint,
  title={“A Hint of it, With Initials”: Adultery, Textuality and Publicity in Jane Austen's Lady Susan},
  author={Russell, Gillian},
  journal={Women's Writing},
  volume={17},
  number={3},
  pages={469--486},
  year={2010},
  publisher={Taylor \& Francis}
}

@article{zouidi2021sleight,
  title={A Sleight of Mind: The Idea of Magic and the Narrative Structure of A Christmas Carol by Charles Dickens},
  author={Zouidi, Nizar},
  journal={Zagadnienia Rodzaj{\'o}w Literackich},
  volume={64},
  number={4},
  pages={35--47},
  year={2021},
  publisher={{\L}{\'o}dzkie Towarzystwo Naukowe}
}

@article{loeve1946fonctions,
  title={Fonctions al{\'e}atoires {\`a} d{\'e}composition orthogonale exponentielle},
  author={Lo{\`e}ve, Michel},
  journal={La Revue Scientifique},
  volume={84},
  pages={159--162},
  year={1946}
}

@article{karhunen1946spektraltheorie,
  title={Zur spektraltheorie stochastischer prozesse},
  author={Karhunen, Kari},
  journal={Ann. Acad. Sci. Fennicae, AI},
  volume={34},
  year={1946}
}

@article{oldham2019consistency,
  title={Consistency and differences between centrality measures across distinct classes of networks},
  author={Oldham, Stuart and Fulcher, Ben and Parkes, Linden and Arnatkevici{\=u}t{\.e}, Aurina and Suo, Chao and Fornito, Alex},
  journal={PloS one},
  volume={14},
  number={7},
  pages={e0220061},
  year={2019},
  publisher={Public Library of Science San Francisco, CA USA}
}

@article{prvzulj2007biological,
  title={Biological network comparison using graphlet degree distribution},
  author={Pr{\v{z}}ulj, Nata{\v{s}}a},
  journal={Bioinformatics},
  volume={23},
  number={2},
  pages={e177--e183},
  year={2007},
  publisher={Oxford University Press}
}

@article{yaverouglu2014revealing,
  title={Revealing the hidden language of complex networks},
  author={Yavero{\u{g}}lu, {\"O}mer Nebil and Malod-Dognin, No{\"e}l and Davis, Darren and Levnajic, Zoran and Janjic, Vuk and Karapandza, Rasa and Stojmirovic, Aleksandar and Pr{\v{z}}ulj, Nata{\v{s}}a},
  journal={Scientific reports},
  volume={4},
  number={1},
  pages={4547},
  year={2014},
  publisher={Nature Publishing Group UK London}
}

@article{jacques2014functional,
  title={Functional data clustering: a survey},
  author={Jacques, Julien and Preda, Cristian},
  journal={Advances in Data Analysis and Classification},
  volume={8},
  number={3},
  pages={231--255},
  year={2014},
  publisher={Springer}
}

@article{higgins2024addressing,
  title={Addressing class imbalance in functional data clustering: C. Higgins, M. Carey},
  author={Higgins, Catherine and Carey, Michelle},
  journal={Advances in Data Analysis and Classification},
  volume={19},
  number={4},
  pages={1023--1050},
  year={2025},
  publisher={Springer}
}

@article{wang2024review,
  title={Review on functional data classification},
  author={Wang, Shuoyang and Huang, Yuan and Cao, Guanqun},
  journal={Wiley Interdisciplinary Reviews: Computational Statistics},
  volume={16},
  number={1},
  pages={e1638},
  year={2024},
  publisher={Wiley Online Library}
}

@article{jain2016statistical,
  title={Statistical graph space analysis},
  author={Jain, Brijnesh J},
  journal={Pattern Recognition},
  volume={60},
  pages={802--812},
  year={2016},
  publisher={Elsevier}
}

@article{ma2020ensemble,
  title={An ensemble of random graphs with identical degree distribution},
  author={Ma, Fei and Wang, Xiaomin and Wang, Ping},
  journal={Chaos: An Interdisciplinary Journal of Nonlinear Science},
  volume={30},
  number={1},
  year={2020},
  publisher={AIP Publishing}
}

@article{cavallaro2024sensitivity,
  title={On the sensitivity of centrality metrics},
  author={Cavallaro, Lucia and De Meo, Pasquale and Fiumara, Giacomo and Liotta, Antonio},
  journal={Plos one},
  volume={19},
  number={5},
  pages={e0299255},
  year={2024},
  publisher={Public Library of Science San Francisco, CA USA}
}

@article{dryden2009non,
  title={Non-Euclidean statistics for covariance matrices, with applications to diffusion tensor imaging},
  author={Dryden, Ian L and Koloydenko, Alexey and Zhou, Diwei},
  journal={The Annals of Applied Statistics},
  pages={1102--1123},
  year={2009},
  publisher={JSTOR}
}

@book{jackson2008social,
  title={Social and economic networks},
  author={Jackson, Matthew O and others},
  volume={3},
  year={2008},
  publisher={Princeton university press Princeton}
}

@book{bollobas1998modern,
  title={Modern graph theory},
  author={Bollob{\'a}s, B{\'e}la},
  volume={184},
  year={1998},
  publisher={Springer Science \& Business Media}
}

@article{zomorodian2005computing,
  title={Computing Persistent Homology},
  author={Zomorodian, Afra and Carlsson, Gunnar},
  journal={Discrete \& Computational Geometry},
  volume={33},
  number={2},
  pages={249},
  year={2005},
  publisher={Springer Nature BV}
}

@article{huang2011temporal,
  title={Temporal dynamics of host molecular responses differentiate symptomatic and asymptomatic influenza a infection},
  author={Huang, Yongsheng and Zaas, Aimee K and Rao, Arvind and Dobigeon, Nicolas and Woolf, Peter J and Veldman, Timothy and {\O}ien, N Christine and McClain, Micah T and Varkey, Jay B and Nicholson, Bradley and others},
  journal={PLoS genetics},
  volume={7},
  number={8},
  pages={e1002234},
  year={2011},
  publisher={Public Library of Science San Francisco, USA}
}

@article{carey2018big,
  title={A big data pipeline: Identifying dynamic gene regulatory networks from time-course Gene Expression Omnibus data with applications to influenza infection},
  author={Carey, Michelle and Ram{\'\i}rez, Juan Camilo and Wu, Shuang and Wu, Hulin},
  journal={Statistical methods in medical research},
  volume={27},
  number={7},
  pages={1930--1955},
  year={2018},
  publisher={SAGE Publications Sage UK: London, England}
}

@article{newman2010networks,
  title={Networks: An Introduction, Oxford University Press},
  author={Newman, M},
  journal={},
  volume={},
  year={2010}
}

@article{munch2015probabilistic,
  title={Probabilistic Fr{\'e}chet means and statistics on vineyards. Electron},
  author={Munch, E and Bendich, P and Turner, K and Mukherjee, S and Mattingly, J and Harer, J},
  journal={J. Statist},
  volume={9},
  number={1},
  pages={1173--1204},
  year={2015}
}

@article{turner2014frechet,
  title={Fr{\'e}chet means for distributions of persistence diagrams},
  author={Turner, Katharine and Mileyko, Yuriy and Mukherjee, Sayan and Harer, John},
  journal={Discrete \& Computational Geometry},
  volume={52},
  pages={44--70},
  year={2014},
  publisher={Springer}
}

@book{networkbook,
  author = {Kolaczyk, E. D},
  publisher = {Springer},
  title = {Statistical Analysis of Network Data
Methods and Models},
  year = 2009
}

@book{ramsay2009functional,
  title={Functional Data Analysis with R and MATLAB},
  author={Ramsay, J. and Hooker, G. and Graves, S.},
  isbn={9780387981857},
  lccn={2009928040},
  series={Use R!},
  year={2009},
  publisher={Springer New York}
}

@book{kokoszka2017introduction,
  title={Introduction to Functional Data Analysis},
  author={Kokoszka, P. and Reimherr, M.},
  isbn={9781498746342},
  lccn={2016045222},
  series={Chapman \& Hall / CRC numerical analysis and scientific computing},
  year={2017},
  publisher={CRC Press}
}

@article{severn2022manifold,
  title={Manifold valued data analysis of samples of networks, with applications in corpus linguistics},
  author={Severn, Katie E and Dryden, Ian L and Preston, Simon P},
  journal={The Annals of Applied Statistics},
  volume={16},
  number={1},
  pages={368--390},
  year={2022},
  publisher={Institute of Mathematical Statistics}
}

@article{ginestet2017hypothesis,
  title={Hypothesis testing for network data in functional neuroimaging},
  author={Ginestet, Cedric E and Li, Jun and Balachandran, Prakash and Rosenberg, Steven and Kolaczyk, Eric D},
  journal={The Annals of Applied Statistics},
  pages={725--750},
  year={2017},
  publisher={JSTOR}
}

@article{kolaczyk2020averages,
  title={Averages of unlabeled networks: Geometric characterization and asymptotic behavior},
  author={Kolaczyk, Eric D and Lin, Lizhen and Rosenberg, Steven and Walters, Jackson and Xu, Jie},
  journal={The Annals of Applied Statistics},
  pages={514--538},
  year={2020},
  
}

@article{guo2021quotient,
  title={A quotient space formulation for generative statistical analysis of graphical data},
  author={Guo, Xiaoyang and Srivastava, Anuj and Sarkar, Sudeep},
  journal={Journal of Mathematical Imaging and Vision},
  volume={63},
  pages={735--752},
  year={2021},
  publisher={Springer}
}

@article{lutgehetmann2020computing,
  title={Computing persistent homology of directed flag complexes},
  author={L{\"u}tgehetmann, Daniel and Govc, Dejan and Smith, Jason P and Levi, Ran},
  journal={Algorithms},
  volume={13},
  number={1},
  pages={19},
  year={2020},
  publisher={MDPI}
}

@article{carlsson2009topology,
  title={Topology and data},
  author={Carlsson, Gunnar},
  journal={Bulletin of the American Mathematical Society},
  volume={46},
  number={2},
  pages={255--308},
  year={2009}
}

@article{edelsbrunner2002topological,
  title={Topological persistence and simplification},
  author={Edelsbrunner and Letscher and Zomorodian},
  journal={Discrete \& Computational Geometry},
  volume={28},
  pages={511--533},
  year={2002},
  publisher={Springer}
}

@article{bubenik2015statistical,
  title={Statistical topological data analysis using persistence landscapes.},
  author={Bubenik, Peter and others},
  journal={J. Mach. Learn. Res.},
  volume={16},
  number={1},
  pages={77--102},
  year={2015}
}

@inproceedings{chazal2014stochastic,
  title={Stochastic convergence of persistence landscapes and silhouettes},
  author={Chazal, Fr{\'e}d{\'e}ric and Fasy, Brittany Terese and Lecci, Fabrizio and Rinaldo, Alessandro and Wasserman, Larry},
  booktitle={Proceedings of the thirtieth annual symposium on Computational geometry},
  pages={474--483},
  year={2014}
}

@article{mileyko2011probability,
  title={Probability measures on the space of persistence diagrams},
  author={Mileyko, Yuriy and Mukherjee, Sayan and Harer, John},
  journal={Inverse Problems},
  volume={27},
  number={12},
  pages={124007},
  year={2011},
  publisher={IOP Publishing}
}

@article{berry2020functional,
  title={Functional summaries of persistence diagrams},
  author={Berry, Eric and Chen, Yen-Chi and Cisewski-Kehe, Jessi and Fasy, Brittany Terese},
  journal={Journal of Applied and Computational Topology},
  volume={4},
  number={2},
  pages={211--262},
  year={2020},
  publisher={Springer}
}

@article{jain2009structure,
  title={Structure Spaces.},
  author={Jain, Brijnesh J and Obermayer, Klaus},
  journal={Journal of Machine Learning Research},
  volume={10},
  number={11},
  year={2009}
}

@article{calissano2023populations,
  title={Populations of Unlabelled Networks: Graph Space Geometry and Generalized Geodesic Principal Components},
  author={Calissano, Anna and Feragen, Aasa and Vantini, Simone},
  journal={Biometrika},
  volume={111},
  number={1},
  pages={147--170},
  year={2024},
  publisher={Oxford University Press}
}

@article{bubenik2017persistence,
  title={A persistence landscapes toolbox for topological statistics},
  author={Bubenik, Peter and D{\l}otko, Pawe{\l}},
  journal={Journal of Symbolic Computation},
  volume={78},
  pages={91--114},
  year={2017},
  publisher={Elsevier}
}

@article{edgar2002gene,
  title={Gene Expression Omnibus: NCBI gene expression and hybridization array data repository},
  author={Edgar, Ron and Domrachev, Michael and Lash, Alex E},
  journal={Nucleic acids research},
  volume={30},
  number={1},
  pages={207--210},
  year={2002},
  publisher={Oxford University Press}
}

@book{edelsbrunner2010computational,
  title={Computational topology: an introduction},
  author={Edelsbrunner, Herbert and Harer, John L},
  year={2010},
  publisher={American Mathematical Society}
}

@article{tauzin2021giotto,
  title={giotto-tda:: A topological data analysis toolkit for machine learning and data exploration},
  author={Tauzin, Guillaume and Lupo, Umberto and Tunstall, Lewis and P{\'e}rez, Julian Burella and Caorsi, Matteo and Medina-Mardones, Anibal M and Dassatti, Alberto and Hess, Kathryn},
  journal={Journal of Machine Learning Research},
  volume={22},
  number={39},
  pages={1--6},
  year={2021}
}

@article{ramsay1991some,
  title={Some tools for functional data analysis},
  author={Ramsay, James O and Dalzell, CJ1125714},
  journal={Journal of the Royal Statistical Society Series B: Statistical Methodology},
  volume={53},
  number={3},
  pages={539--561},
  year={1991},
  publisher={Oxford University Press}
}

@article{erdos1959random,
  title={On random graphs I},
  author={Erdös, P. and Rényi, A},
  journal={Publicationes Mathematicae Debrecen},
  volume={6},
  number={290-297},
  year={1959}
}

@article{mahlberg2016clic,
  title={CLiC Dickens: Novel uses of concordances for the integration of corpus stylistics and cognitive poetics},
  author={Mahlberg, Michaela and Stockwell, Peter and Joode, Johan de and Smith, Catherine and O'Donnell, Matthew Brook},
  journal={Corpora},
  volume={11},
  number={3},
  pages={433--463},
  year={2016},
  publisher={Edinburgh University Press 22 George Square, Edinburgh EH8 9LF UK}
}

\end{document}